\RequirePackage{lineno} 
\documentclass[preprint2]{proto}
\usepackage{natbib}

\usepackage{color}
 
\bibpunct{(}{)}{;}{a}{,}{,}
\usepackage{times}

\def \mum{$\mu$m}
\newcommand \kms{km~$\rm{s}^{-1}$}

\newcommand \lsol{L$_{\odot}$}
\newcommand \msol{M$_{\odot}$}
\newcommand \msolyr{M$_{\odot}$~yr$^{-1}$}

\newfont{\rten}{cmr10} 

\newcommand{\hii}{H{\sc ii}}

\newcommand{\HII}{H{\sc ii}}
\newcommand{\HI}{H{\sc i}}
\newcommand{\Hi}{H{\sc i}}
\newcommand{\hi}{H{\sc i}}

\def \arcmin{\hbox{$^\prime$}}
\def \arcsec{\hbox{$^{\prime\prime}$}}
\renewcommand{\deg}{$^{\circ}$}

\begin{document}

\setpagewiselinenumbers
\switchlinenumbers 
\renewcommand\linenumberfont{\normalfont\small}
\modulolinenumbers[5]
%

\title{\textbf{\LARGE The Milky Way as a Star Formation Engine}}


\author {\textbf{\large $^1$Sergio Molinari},
               \textbf{\large $^2$John Bally}, 
               \textbf{\large $^3$Simon Glover},
               \textbf{\large $^4$Toby Moore}, 
               \textbf{\large $^5$Alberto Noriega-Crespo},   
               \textbf{\large $^6$Ren\'e Plume},
               \textbf{\large $^{7,8,9}$Leonardo Testi},
               \textbf{\large $^{10}$Enrique V\'azquez-Semadeni}, 
               \textbf{\large $^{11}$Annie Zavagno},
              \textbf{\large $^{12}$Jean-Philippe Bernard}, 
               \textbf{\large $^{13}$Peter Martin}}
\affil{\small\em 
{\bf 1.}  
INAF-Istituto di Astrofisica e Planetologia Spaziali, Via Fosso del Cavaliere 100, I-00133 Rome, Italy, 
{\bf 2.}
Center for Astrophysics and Space Astronomy, UCB 389, University of Colorado at Boulder, Boulder CO 80389, USA 
{\bf 3.} 
Universit\"at Heidelberg, Zentrum f\"ur Astronomie, Institut f\"ur Theoretische Astrophysik, Albert-Ueberle-Str.\ 2, 69120 Heidelberg, Germany, 
{\bf 4.} 
Astrophysics Research Institute, Liverpool John Moores University, Twelve Quays House, Egerton Wharf, Birkenhead CH41 1LD, UK, 
{\bf 5.} 
Space Telescope Science Institute, 3700 San Martin Drive, Baltimore, MD 21218, USA 
{\bf 6.} 
University of Calgary, Dept. of Physics \& Astronomy, 2500 University Dr. NW, Calgary, AB, T2N1N4, Canada, 
{\bf 7.} 
ESO, Karl Schwarzschild str. 2, D-85748 Garching, Germany, 
{\bf 8.} 
INAF--Osservatorio Astrofisico di Arcetri, Largo E. Fermi 5, I-50125 Firenze, Italy
{\bf 9.} 
Excellence Cluster Universe, Boltzmannstr. 2, D-85748 Garching, Germany
{\bf 10.} 
Centro de Radioastronomia y Astrofisica, Universidad Nacional Autonoma de Mexico, Campus Morelia, Apdo. Postal 3-72, Morelia, 58089, Mexico, 
{\bf 11.} 
Aix Marseille Universit\'e, CNRS, LAM (Laboratoire d'Astrophysique de Marseille) UMR 7326, 13388, Marseille, France 
{\bf 12.} 
Institut de Recherche en Astrophysique et Plan\' etologie, 9, av. du Colonel-Roche - BP 44 346, 31028 Toulouse Cedex 4  France, 
{\bf 13.} 
Canadian Institute for Theoretical Astrophysics, University of Toronto, 60 St. George Street, Toronto, ON M5S 3H8, Canada
}

\begin{abstract}
\baselineskip = 11pt
\leftskip = 0.65in 
\rightskip = 0.65in
\parindent=1pc
{\small 

The cycling of material from the interstellar medium (ISM) into stars
and the return of stellar ejecta into the ISM is the engine that
drives the {\it galactic ecology} in normal spirals.  This ecology is
a cornerstone in the formation and evolution of galaxies through
cosmic time.  
There remain major observational and theoretical challenges in
determining the processes responsible for 
converting the low-density, diffuse components of the ISM into dense
molecular clouds,
forming  dense filaments and clumps,
fragmenting them into stars, expanding OB associations, and bound
clusters,
and characterizing the feedback that limits the rate and efficiency
of star formation.
This formidable task can be attacked effectively for the first time
thanks to the synergistic combination of new global-scale surveys of
the Milky Way from infrared to radio wavelengths, offering the
possibility of bridging the gap between local and extragalactic star
formation studies.

The Herschel Hi-GAL survey, with its five-band 70--500\mum\ full
Galactic Plane mapping at 6--36\arcsec\ resolution, is the keystone of
a set of continuum surveys that include GLIMPSE(360)+MIPSGAL@Spitzer,
WISE and MSX, ATLASGAL@APEX, BGPS@CSO, and CORNISH@VLA. This suite
enables us to measure the Galactic distribution and physical
properties of dust on all scales and in all components of the ISM from
diffuse clouds to filamentary complexes and hundreds of thousands of
dense clumps.
A complementary suite of spectroscopic surveys in various atomic and
molecular tracers is providing the chemical fingerprinting of dense
clumps and filaments, as well as essential kinematic information to
derive distances and thus transform panoramic data into a 3D
representation.

The latest results emerging from these Galaxy-scale surveys are
reviewed.  New insights into cloud formation and evolution, filaments
and their relationship to channeling gas onto gravitationally-bound
clumps, the properties of these clumps, density thresholds for
gravitational collapse, and star and cluster formation rates are
discussed.
\\~\\~\\~}   

\end{abstract}  


\section{Introduction}

Phase changes in the Galactic ISM are to a large extent controlled by
the formation of massive stars.  The cycling of the ISM from mostly
neutral atomic (\HI) clouds into molecular (H$_2$) clouds, traced by
low-excitation species such as OH and CO, leads to the formation of
dense, self-gravitating clumps and cores traced by high-density
species such as NH$_3$, CS, HCN, HCO$^+$, and N$_2$H$^+$, and by other
even higher dipole moment molecules in regions where stars form.

Dust continuum emission in the mid-IR, far-IR, sub-millimeter (submm),
and millimeter ranges of the spectrum reveal progressively cooler and
higher column density dust associated with all phases of the ISM, and
with the high-density material in protostellar envelopes and disks.
Dust extinction measurements in the UV, visual, near-IR, and mid-IR
enable detection of progressively higher column densities of dust
located in front of background stellar and diffuse emission sources.

Feedback from young stars limits the rate and efficiency of star
formation by generating turbulence and disrupting the parent clouds.
In giant molecular clouds (GMCs) feedback, usually dominated by the
most massive young stellar objects (YSOs), tends to disrupt the parent
cloud by the time 2--20\%\ of its mass has formed stars.  Ionization
and shocks produced by massive stars, associations, and clusters
convert the remaining gas into the $10^4$--$10^8$~K photoionized and
shock-heated phases of the ISM -- \HII\ regions traced by H and He
recombination lines and free-free emission, hot superbubbles traced by
X-ray emission, and 0.1 to 1 kpc-scale supershells -- that eventually
cool, condense, and reform the cool $\sim$ $20$ to $10^4$~K \HI\
phase.  Compression by shocks and gravity leads to the formation of
new GMCs.

In the Solar vicinity, atoms cycle through this loop on a time scale
of 50--100~My.  A mean star formation efficiency (SFE) of 5\% per GMC
implies that atoms on average pass through this loop about 20 times
before being incorporated into a star.  This {\it galactic ecology} is
modulated by large-scale processes such as spiral arms and the central
bar of the Milky Way.  While star formation depletes the ISM at a rate
of $\sim$ 2 \msol~y$^{-1}$, infall of gas from the Local Group
supplements it at a highly uncertain rate of 0.1 to 1 \msol~y$^{-1}$.
The balance between star formation, which sequesters matter for the
main-sequence lifetime of stars, and recycling from stellar winds and
dying stars, supplemented by infall from the Local Group, determines
the time scale on which the Galactic ISM is depleted.  Large-scale
Galactic plane surveys of the last decade provide the data required to
flesh out the details of this {\it galactic ecology}.

The Milky Way, a mildly-barred, gas-rich spiral galaxy, supports the
most active star formation in the Local Group.  The Galactic disk
contains about 1--$3 \times 10^9$ M$_{\odot}$ of H$_2$ and about 2--$6
\times 10^9$ M$_{\odot}$ of \HI\
\citep{Combes1991,KalberlaKerp2009}. The molecular disk becomes
prominent from Galactocentric radius $R_{gal} \sim$ 3 kpc, beyond the
central bar, with surface density peaking at about $R_{gal} \sim$ 4--6
kpc (often referred to as the {\it Molecular Ring}).  It declines
exponentially toward larger radii but can be traced well beyond
$R_{gal} \sim $10 kpc.
The \HI\ disk extends to beyond $R_{gal} \sim$ 20 kpc.
The disk ISM is dominated by a four-arm spiral pattern consisting of
two major and two minor spiral arms and a variety of inter-arm spurs.
The Sun is currently located in an inter-arm region between the
Sagittarius and Perseus spiral arms, near a spur which extends from a
distance of several kpc in the direction of Cygnus to at least 1 kpc
beyond Orion in the opposite direction \citep{Xu2013}.
In molecular tracers such as CO, the arm--inter-arm contrast in the
Molecular Ring is around 3:1, but in the outer Galaxy beyond the Solar
circle at $R_{gal} \approx$ 8.5 kpc the contrast is much higher,
approaching a value of 40:1 toward the Perseus arm.

Although Galactic rotation is well described by circular motions with
orbit speeds ranging from 200 to 250 km~s$^{-1}$ from $R_{gal} \sim$ 1
kpc to beyond 20 kpc, substantial radial motions are seen in gas
tracers toward the Galactic Center and Anti-center.  The line of sight
toward the Galactic Center shows three clearly-defined spiral arms,
distinguished by their negative radial velocities with respect to the
local standard of rest (LSR; defined by the mean motion of stars in
the Solar vicinity).  The innermost is the so-called 3-kpc arm in the
Molecular Ring ($V_{LSR} \approx -60$ km~s$^{-1}$), next is the Scutum
arm ($V_{LSR} \approx -35$ km~s$^{-1}$), and then closest ($\sim$ 2
kpc) is the Sagittarius arm ($V_{LSR} \approx -15$ km~s$^{-1}$).
Radial streaming motions of order 10 to 30 km~s$^{-1}$ are also seen
toward the Perseus arm and the far outer arm beyond that.  These
motions might reflect the gravitational potential well depth of the
spiral arms.

The inner edge of the Molecular Ring at $R_{gal} \sim 3$~kpc lies near
the outer radius of the Galactic bulge and bar.  The central 3-kpc
region of the Galaxy is dominated by the stellar bar whose major axis
is inclined by 20\deg\ to 40\deg\ with respect to our line of sight
\citep{binney91,MorrisSerabyn1996}.  Within 0.5 kpc of the nucleus
lies the Central Molecular Zone (CMZ) containing $\sim 10$\% of the
molecular gas in the Galaxy.  The CMZ clouds are one to two orders of
magnitude denser than GMCs in the Molecular Ring, and an order of
magnitude more turbulent.

\section{Physical Properties of the Milky Way ISM from Dust and Gas Tracers}
\label{par_surveys}


With the advent of photography and multi-color photometry in the late
1800s and early 1900s the presence of interstellar dust was recognized
by its extinction and reddening effect on the light from distant
stars.  Evidence for interstellar gas followed in the 1930s with the
detection of narrow absorption lines of sodium and potassium and a few
simple molecules such as CH and CH$^+$.  The discovery of the 21 cm
line of \HI\ in the 1940s led to the first all-sky surveys of the ISM
and the recognition that the Galaxy contains more than $10^{9}$ \msol\
of gas.  The detection of absorption and thermal emission from OH,
NH$_3$, CO, and many other molecules in the 1960s and 1970s led to the
discovery of molecular clouds and the recognition that GMCs were the
birth sites of stars.

The last 30 years have seen an impressive set of infrared, millimeter,
and radio surveys both in spectroscopy and in continuum bands that
have covered the Galactic Plane.  Table~\ref{surveys} gives a
representative list.  These have increased exponentially the amount of
information available but at the same time have exposed the complexity
of the puzzle of the {\it galactic ecology}.

\begin{table*}[t]
\caption{List of most representative surveys covering the Galactic Plane}
\begin{flushleft}
\begin{tabular}{lcl}\hline\hline
Surveys facilities & $\lambda$ or lines & Surveys notes \\ \hline
\multicolumn{3}{c}{Ground-based} \\ \hline
\parbox{2.5cm}{Columbia/CfA} & \parbox{4.5cm}{CO,  $^{13}$CO} & \parbox{8.5cm}{9 - 25\arcmin\ resolution \citep{dame2001}} \\
\parbox{2.5cm}{DRAO/ATCA/VLA \\ \\ \\ \\} & \parbox{4.5cm}{HI-21 cm OH/H$\alpha$-RRL/1-2GHz cont. 5GHz cont.\\ \\ \\ } & \parbox{8.5cm}{IGPS: unbiased HI-21cm  255\deg$\leq l \leq 357$\deg\ and 18\deg$\leq l \leq 147$\deg\ \citep{McClure+2001,gib00,Stil+2006} + THOR: unbiased HI-21cm/OH/H$\alpha$-RRLs/1-2GHz cont. 15\deg$\leq l \leq 67$\deg\ (Beuther et al. in prep.)+ CORNISH: 5GHz continuum 10\deg$\leq l \leq 65$\deg\ \citep{Hoare2012}} \\
\parbox{2.5cm}{FCRAO 14 m \\ }  & \parbox{4.5cm}{CO,  $^{13}$CO \\ } & \parbox{8.5cm}{55\arcsec\ resolution. Galactic Ring Survey \citep{Jackson+2006} + Outer Galaxy Survey \citep{Heyer1998}} \\
\parbox{2.5cm}{Mopra 22 m \\ \\ \\ \\ }  & \parbox{4.5cm}{CO,
  $^{13}$CO, N$_2$H$^+$, (NH$_3$ + H$_2$O) maser,
  HCO$^+$/H$^{13}$CO$^+$ + others \\ \\ } & \parbox{8.5cm}{HOPS:
  \citep{Walsh2011,Purcell_HOPS_2012}, MALT90: $\sim$ 2000 clumps 20\deg\ $\geq l \geq -60$\deg\ \citep{MALT90_2013}, Southern GPS CO: unbiased 305\deg$\leq l \leq 345$\deg\ \citep{Burton+2013}, ThrUMMS: unbiased 300\deg$\leq l \leq 358$\deg\ \citep{Barnes+2013}, CMZ: \citep{Jones2012, Jones+2013}} \\
\parbox{2.5cm}{Parkes} & \parbox{4.5cm}{CH$_3$OH maser} & \parbox{8.5cm}{Methanol MultiBeam Survey \citep{Green+2009}} \\ 
\parbox{2.5cm}{NANTEN/ NANTEN2}  & \parbox{4.5cm}{CO,  $^{13}$CO, C$^{18}$O \\ } & \parbox{8.5cm}{NGPS: unbiased, 200\deg\ $ \leq l \leq 60$\deg\ \citep{Mizuno+2004} + NASCO: unbiased in progress, 160\deg\ $ \leq l \leq 80$\deg} \\
\parbox{2.5cm}{CSO 10 m \\ } & \parbox{4.5cm}{1.3 mm continuum \\ } & \parbox{8.5cm}{Bolocam Galactic Plane Survey (BGPS), 33\arcsec\ \citep{Aguirre2011}} \\
\parbox{2.5cm}{APEX 12 m} & \parbox{4.5cm}{870 \mum\ continuum} & \parbox{8.5cm}{ATLASGAL, 60\deg $\geq l \geq -80$\deg\ \citep{Schuller2009}} \\ \hline
\multicolumn{3}{c}{Space-borne} \\ \hline
\parbox{2.5cm}{IRAS} & \parbox{4.5cm}{12, 25, 60 and 100 \mum\ cont.} & \parbox{8.5cm}{3-5\arcmin, 96\% of the sky} \\
  \parbox{2.5cm}{MSX} & \parbox{4.5cm}{8.3, 12.1, 14.7, 21.3 \mum\ cont.} & \parbox{8.5cm}{Full Galactic Plane \citep{Price+2001}} \\
  \parbox{2.5cm}{WISE}  & \parbox{4.5cm}{3.4, 4.6, 11, 22 \mum\ continuum} &  \parbox{8.5cm}{All-sky \citep{Wright+2010}} \\
  \parbox{2.5cm}{Akari}	  & \parbox{4.5cm}{65, 90, 140, 160 \mum\ continuum} &	\parbox{8.5cm}{All-sky \citep{Ishihara+2010}} \\
  \parbox{2.5cm}{Spitzer \\ \\ }  & \parbox{4.5cm}{3.6, 4.5, 6, 8, 24
    \mum\ continuum \\ \\ }	& \parbox{8.5cm}{GLIMPSE+GLIMPSE360:
    Full Galactic Plane \citep{Benjamin+2003}, \citep{Benjamin+2013} + MIPSGAL, 63\deg $\geq l \geq -62$\deg\ \citep{Carey+2009}}\\
  \parbox{2.5cm}{Planck \\ }  & \parbox{4.5cm}{350, 550, 850, 1382,
    2098, 3000, 4285, 6820, $10^{4}$ \mum\ cont.}
  & \parbox{8.5cm}{All-sky, resolution $\geq$5\arcmin\  \citep{planck2013} \\ } \\
  \parbox{2.5cm}{Herschel} & \parbox{4.5cm}{70, 160, 250, 350, 500 \mum\ cont.} & \parbox{8.5cm}{Hi-GAL: Full Galactic Plane \citep{Molinari2010b}} \\ \hline
\end{tabular}
\end{flushleft}
\label{surveys}
\end{table*}

\subsection{Spectral Line Surveys}
\label{linesurveys}

Detector technology has to a large extent dictated the development of
Galactic plane surveys.  From the late 1970s until the late 1990s,
heterodyne receivers dominated, with a number of evermore detailed
spectral line surveys of the entire sky in \HI, portions of the
Galactic plane in CO, and selected regions such as Orion, Sgr B2, and
other star forming complexes as well as a few post-main sequence stars
in a range of other molecular species.

Since the early 1970s, CO emission has been the most commonly used
tracer of molecular gas in the Galaxy.  The CO molecule is the most
abundant species in molecular gas next to H$_2$ and He.  Its small
($\sim$ 0.1 Debye) dipole moment results in bright CO emission
widespread along the Galactic plane.  Its large abundance ($\sim
10^{-4}$ times that of H$_2$), high opacity, and consequent radiative
trapping enables CO to trace molecular gas at H$_2$ densities above
$\sim 10^2$~cm$^{-3}$.  The first comprehensive survey in the CO $J =
1-0$ line \citep{dame1987} covered the Galactic plane over a latitude
range of $10^\circ - 20^\circ$ with a grid resolution of
$0.5^\circ$, and parts of Galactic plane with a resolution of
$9$\arcmin, revealing the large-scale spatial and velocity
distribution of CO-emitting molecular gas.
\cite{dame2001} combined spectra from 31 sub-surveys to cover $4^\circ
- 10^\circ$ in latitude with $9$\arcmin\ to $15$\arcmin\ angular
resolution.  
The first all-sky image of CO emission was produced from Planck data
through careful analysis detector by detector of Planck HFI continuum
images \citep{Planck_CO_2013}.  Velocity-integrated maps of the three
lowest rotational transitions of CO were produced at $\sim 5$\arcmin\
resolution.  In the regions of overlap, the agreement between the Dame
et al.\ and the Planck products is excellent.  The Planck CO images
reveal many new high-latitude CO-emitting clouds.

Higher angular resolution ($<$ few arcminute) CO surveys cover
selected portions of the Galactic Plane.  The Five College Radio
Astronomy Observatory (FCRAO) 14~m telescope surveyed the Northern
Galactic plane with a resolution of $\sim$45\arcsec\ \citep{Heyer1998,
  RomanDuval2010}.  Data covering the remaining portions of the Northern
Galactic plane were obtained before FCRAO was closed, but have not yet
been published.  Selected regions of the Southern sky have been
obtained with the NANTEN2, ASTE, APEX, and Mopra telescopes.  Although
CO is an excellent tracer of H$_{2}$, most of the gas traced by this
low-excitation species is not directly associated with star formation.

Detailed studies of selected nearby molecular clouds such as those in
Taurus, Ophiuchus, Perseus, and Orion, as well as of clouds in more
distant portions of the Galactic plane, demonstrated that high-dipole
moment ($\sim$2 to 4 Debye) species such as NH$_3$, CS, HCN, HCO$^+$,
and N$_2$H$^+$ are more closely associated with dense,
self-gravitating clumps and cores associated with the formation of
clusters and individual stars.  Recently, the Mopra telescope surveyed
the Southern Milky Way in the 1.3 cm NH$_3$ and H$_2$O maser lines
(HOPS; \citealt{Walsh2011}).  Mopra is also conducting surveys of
dense gas tracers around 3 mm (MALT90;
\citealt{MALT90_2011,MALT90_2013}), CH, and CO isotopologues (CHaMP;
\citealt{Barnes2011}).

\subsection{Continuum Surveys}
\label{continuumsurveys}

The IRAS survey in the mid-1990s provided the first far-IR all-sky
survey in the dust continuum.  The development of focal plane arrays
in the IR to far-IR wavelength region and submm bolometer arrays in
the late 1990s led to a series of surveys including ISO, MSX, and COBE
from space.  The last decade has seen a number of large continuum
surveys with Spitzer, AKARI, WISE, and, most recently, Herschel and
Planck.  The 12 m APEX telescope was used for the ATLASGAL survey of
the Southern Galactic plane at 870 $\mu$m \citep{Schuller2009}.  The
10 m CSO was used to obtain the 1.1 mm Bolocam Galactic Plane Survey
(BGPS) of the Northern Galactic plane at 1300 $\mu$m with a 33\arcsec\
beam \citep{Aguirre2011}.  The 15 m JCMT is being used for 850 and
450$\mu$m SCUBA2 surveys of the Galactic plane and several nearby
Gould Belt clouds.

A major breakthrough in the study of massive star and cluster
formation was the discovery of infrared dark clouds (IRDCs), which
trace the highest column density clumps and cores in GMCs.  Unbiased
extractions of IRDCs from the MSX maps produced catalogs of several
thousand clouds \citep{Simon+2006}.  Variations in the mid-IR
background, however, can mimic an IRDC without dense and cold
intervening material.  When a more refined approach was used to
extract IRDCs from the higher quality Spitzer/IRAC images of the
GLIMPSE survey \citep{Benjamin+2003}, \cite{Peretto+Fuller2009} found
twice as many clouds as Simon et~al., yet did not confirm many of
Simon et al.'s IRDCs.  It has been suggested that IRDCs are the
high-mass analogues of low-mass prestellar cores.  Early radio
spectroscopic follow-up \citep[e.g.,][]{Pillai+2006} suggested that
star formation in IRDCs is ongoing.  \cite{Peretto+Fuller2009} showed
that as many as 70\% of their IRDCs were indeed associated with a
24-\mum\ compact source from the Spitzer/MIPSGAL Survey
\citep{Carey+2009}, again indicating ongoing star formation.

Dust continuum emission at far-IR, submm, and mm wavelengths provides
a robust tracer of the dust column density and mean grain temperature
along the line of sight, independent of the phase of the associated
ISM.  Surveys with Spitzer and WISE have mapped most of the Galactic
plane at 3.6, 4.5, 6, 8, and 24 $\mu$m with resolutions of a few to
10\arcsec.
The Herschel Space Observatory Galactic Plane Survey (Hi-GAL) and its
successor extensions \citep{Molinari10} have mapped the entire
Galactic plane in five bands at 70, 160, 250, 350, and 500 $\mu$m with
angular resolutions of about 6\arcsec\ to 36\arcsec .

Distances must be determined to extract physical quantities from
continuum data.  Some dust continuum morphological features are
associated with objects for which distances estimates exist.  However,
for the majority of continuum sources there are no such associations.
Continuum surveys of the Galactic plane must be combined with
complementary heterodyne surveys that provide measurements of the
radial velocity of gas associated with dust continuum emission or
absorption features.  Most current distance-determination methods rely
on kinematic distances deduced from an axisymmetric Galactic rotation
model combined with a variety of methods to resolve the kinematic
distance ambiguity (KDA) for objects located within the Solar circle.
However, systematic streaming motions and local peculiar velocities
limit the precision of such distance estimates \citep{Xu2013}.

Radial velocity measurements can be used to associate clumps and
features for which there are no direct distance measurements with
sources for which distance estimates exist.  The Hi-GAL group working on
distances \citep{russeil:2011} used the radial velocities of dust
clumps and linked them to nearby sources (\HII\ regions, stellar
groups, or star forming regions) for which reliable distance estimates
exist.  This ``bootstrapping" approach, together with the use of
near-IR extinction mapping information to help resolve the KDA, has
provided approximate distance estimates for tens of thousands of
Hi-GAL sources \citep{Elia+2014b}.

The key challenge for kinematic distance estimates is resolution of
the KDA.  Distance from the Galactic mid-plane, the presence of an
associated IRDC, a 21 cm ``\HI\ narrow self-absorption" (HINSA)
feature at the LSR velocity of the dust clump (produced by cold atomic
hydrogen in and around the molecular cloud), and the density of
foreground stars can be used to resolve the KDA.  Association with a
high-contrast IRDC, the presence of HINSA, significant displacement from
the Galactic mid-plane, and a low density of foreground stars in the
near-IR favor placement at the near kinematic distance.  Absence of an
associated IRDC and HINSA (because the 21 cm absorption is filled-in
by intervening warm \HI\ emission), along with a high density of
foreground stars in 2MASS, Spitzer, or WISE images, and location close
to the mid-plane favor the far kinematic distance.

\cite{Ellsworth-Bowers2013} have developed an automated statistical
distance determination tool that ingests the Galactic latitude, the
radial velocity, and Spitzer 8 $\mu$m images of the region, and uses
an a-priori model of Galactic diffuse 8 $\mu$m emission to resolve the
KDA using ``Distance-Probability Distribution Functions."  This method
can be generalized to incorporate star counts and reddening toward
dust emission features and the presence or absence of HINSA to provide
a product of probabilities for resolving the KDA.

Higher-precision distance assignments will require precision parallax
measurements toward associated sources, determination of radial
streaming motions, and a full 3D vector model of Galactic rotation.
In a few years, visual-wavelength parallax determinations with the
GAIA mission will provide distances to sources and so might help with
distance determination in relatively nearby regions.  ``The Bar and
Spiral Structure Legacy (BeSSeL) survey" \citep{Bessel2011} will
provide direct parallax determinations toward about 400 Galactic
sources exhibiting cm-wavelength maser emission, with a precision of
$\sim$ 3\% to 10\%.  BeSSeL will provide direct trigonometric
distances throughout the Galaxy because, unlike the GAIA measurements
of stars, radio masers are not subject to extinction.

\section{Overview of  GMCs on the Large Scale}

\begin{figure*}[ht]
\includegraphics[width=1.0\textwidth]{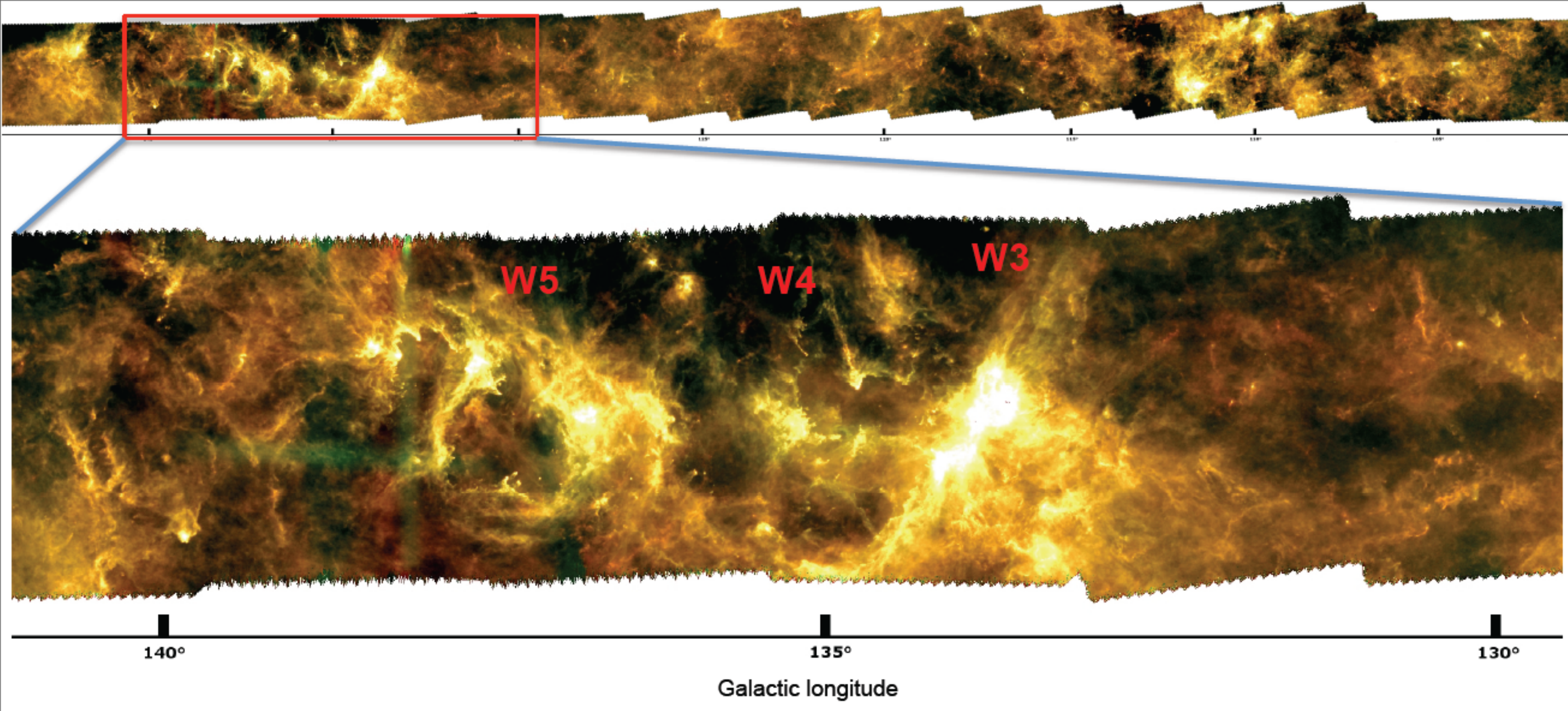}
\caption{Top: Hi-GAL three-color composite of a portion of the Perseus
Arm from data at 70 $\mu$m (blue), 160 $\mu$m (green), and 350 $\mu$m
(red). Bottom: The W5, W4, and W3 GMA and giant HII region complex near $l =
134$\deg .  The red box outlines this region in the top
panel.
%
}
\label{fig:W345}
\end{figure*}



The mass distribution of GMCs can be described by a power-law
$dN / dM \propto M^{\alpha}$ with a cutoff at about $10^{6.5}$ \msol\
for the Galactic disk \citep{Rosolowsky2005}. Because $\alpha = -1.5
\pm 0.1$, massive clouds dominate the total mass of H$_{2}$.

 
GMCs exhibit supersonic internal motions on scales larger than
individual star forming cores ($\sim 0.01$ to $0.05$ pc).  At $R_{gal}
>$ 1~kpc,
tracers such as CO reveal line-widths of order 1 to 10 \kms\ in gas of
mean density $10^2 - 10^3$~cm$^{-3}$ \citep{RomanDuval2010}.  Dense
clumps and cores with $n(H_2) > 10^4$ cm$^{-3}$, traced by high
dipole moment dense gas tracers, typically cover only a few percent of
the projected surface area of GMCs traced by CO emission.
However, in the CMZ such dense gas tracers have line widths of 10 to
over 30 \kms\ and cover most of the projected surface area traced by
CO, indicating that CMZ clouds have one to two orders of magnitude
larger mean densities \citep[e.g.,][]{MorrisSerabyn1996}.

\citet{Heyer2009}, using FCRAO $^{13}$CO data, 
found a dynamic range of nearly two orders of magnitude in the surface
density of the GMCs (a few 10 \msol ~pc$^{-2}$ to a few 1000~\msol
~pc$^{-2}$).
Their generalization of the classic \citep{Larson1981} linewidth-size
scaling relation is $\sigma(\ell) \propto \Sigma^{1/2} \ell^{1/2}$.
%
%
Massive-star forming clumps 
also appear to
follow this relation
\citep{Ballesteros2011a}.  

Molecular clouds have internal filamentary structure, first seen in
deep photographs of nearby dark clouds \citep{Barnard1907} and
extensively explored in CO maps, e.g., in Orion and Taurus
\citep{Bally1987, Narayanan2008, Kirk_Taururs_2013}.  Widespread
filamentary IRDCs, seen in silhouette against
the bright mid-IR background of the Galactic Plane \citep{Egan+1998},
are of heightened interest because IRDCs are the densest, highest
column density parts of GMCs most closely associated with star
formation \citep{Carey+2000}.  Now, thanks to large-scale dust
emission surveys with Herschel, it has become apparent that filaments
are a ubiquitous feature of the ISM at all size scales.  
For example, Figures~\ref{fig:W345} and \ref{fig:NGC7538} show
Herschel Hi-GAL images of filamentary structure in the W3--5 and $l$ =
111 complexes, respectively.

\subsection{Formation}



Theories of GMC formation have evolved enormously since the discovery
of CO. 
Many of the key considerations are discussed
in the review by \citet{McKee2007}.  
 The article by Dobbs et al.\ in this volume 
also covers the formation of molecular clouds and so here we will
simply highlight a few
recent developments.

\begin{figure}[h]
\includegraphics[width=0.485\textwidth]{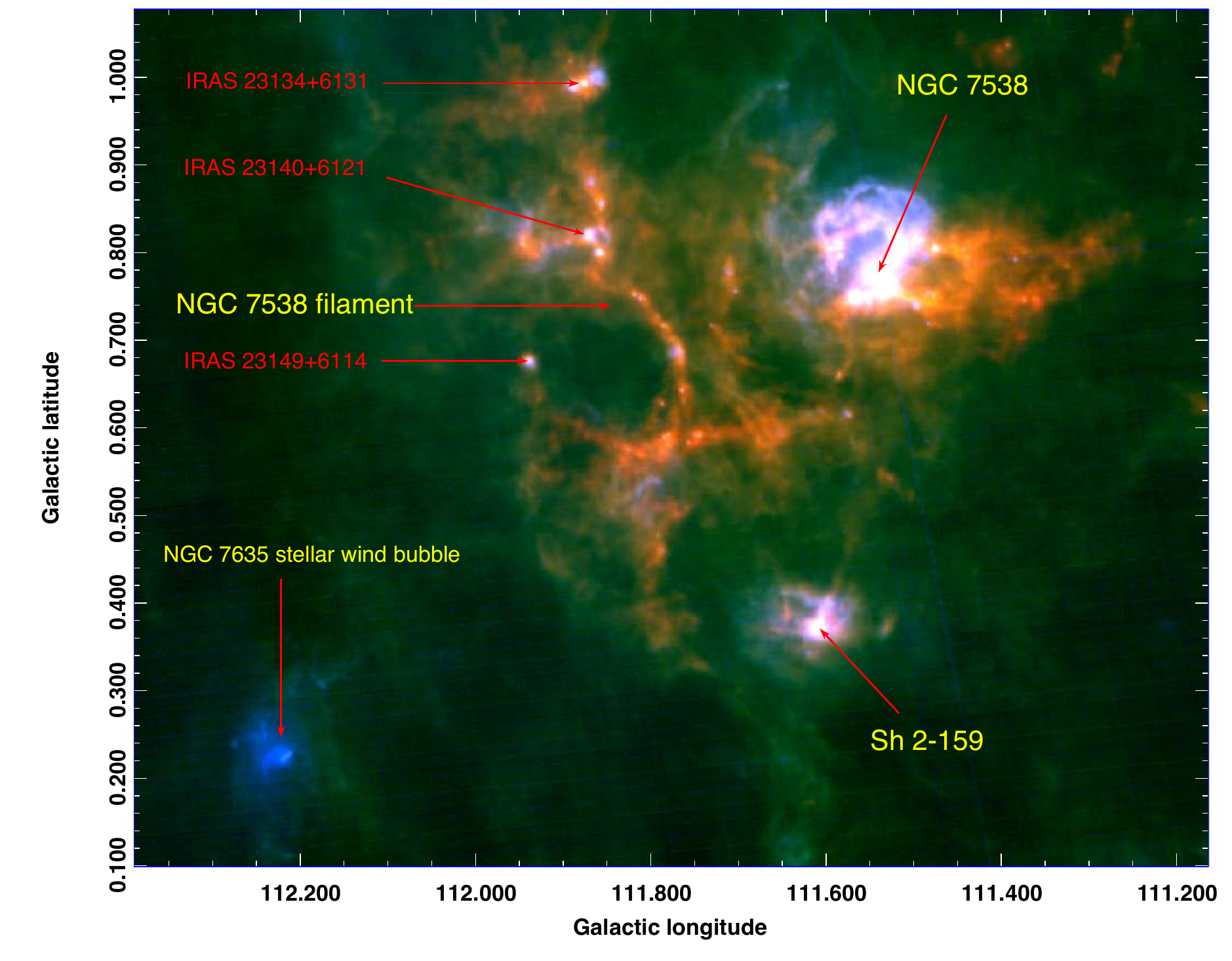}
\caption{ Hi-GAL three-color composite of the GMC complex associated
  with the NGC 7538 HII region at $l$ = 111 in the Perseus Arm, from
  data at 70 $\mu$m (blue), 160 $\mu$m (green), and 350 $\mu$m (red),
  in Galactic coordinates.  }
\label{fig:NGC7538}
\end{figure}

There is growing evidence that GMCs, and the star-forming clumps
within them, form in a multi-step process.
The first step is the concentration of the diffuse non-molecular ISM,
with a mean density of 1~cm$^{-3}$, into \HI\ superclouds and related
Giant Molecular Associations (GMAs) with mean densities of 
10~cm$^{-3}$ \citep{Elmegreen+Elmegreen1987}.  GMAs are characterized
by $10^6 - 4\times 10^7 M_\odot $ \HI\ halos surrounding GMC complexes
at 1 to 2 kpc intervals along the major spiral arms of the Galaxy
\citep{DameElmegreen1986}.  
For example, 
%
in the Perseus arm of the outer Milky Way, GMAs are found at
approximately 2 kpc intervals at longitudes $l \sim 110$\deg\ to
$112$\deg\ associated with NGC 7538 \citep{Fallscheer2013},
$l \sim 132$\deg\ to $136$\deg\
associated with the W3 \citep{Rivera2013}, 
W4, and W5 giant \HII\ regions, and $l \sim
172$\deg\ to $180$\deg\ associated with the Sh2-235 \HII\ region.
Multi-wavelength Herschel imaging brings out the spectacular structures
in such complexes, as illustrated in Figures~\ref{fig:W345} (W3--W5)
and \ref{fig:NGC7538} (NGC 7538).
The GMAs form by the combined effects of self gravity, the
gravitational potential of the Galactic plane, and compression by
spiral arms \citep{Elmegreen1994}; they are
gravitationally bound with internal motions dominated by
turbulence \citep{DameElmegreen1986}.

The subsequent steps concern the development of structures of higher
density.  The second is the formation of dense sheets in post-shock
layers where supersonic turbulence, or the shells and bubbles produced
by massive-star feedback, produce converging and colliding flows
\citep[e.g.,][]{Audit2005}. 
The third step is the
cooling of the post-shock layers and conversion of predominantly
atomic gas into molecular gas.  Fourth is the formation of dense
filaments from sheets.  Fifth is the fragmentation of filaments and
sheets into star-forming clumps.  Recent theoretical developments and observational evidence concerning filaments and dense clumps formation are discussed in \S\ref{filaments} and \S\ref{par_clumps}.
At each step, the density is
increased by about an order of magnitude, thus reaching $n(H) \sim
10^6$ cm$^{-3}$ in clumps and cores. 

It is not yet clear at which step the resulting substructure becomes
recognizable as CO-bright molecular clouds.  As discussed below,
gravity and supersonic convergence can compress the CO-dark H$_2$ to
sufficient density to produce and excite CO, producing CO-bright
structures only shortly before the onset of star formation.



 \subsubsection{CO-free H$_2$;  Dark gas as Fuel for GMC Formation}


Aperture synthesis observations of nearby spiral galaxies
\citep{Allen2004, Heiner2009} comparing their CO and \HII\ region
distributions and radial velocities show that most of the gas traced
by the 21~cm emission line is located downstream from spiral arms,
active sites of star formation, and \HII\ regions.  These results
provide evidence that this \HI\ might be primarily a photodissociation
product.
In this picture, GMCs would primarily form from \HI\ clouds
\emph{through}
CO-dark molecular gas \emph{rather than directly} from \HI\ clouds.

\citet{Pringle2001} hypothesized a component of the ISM in which
hydrogen is mostly molecular but that is CO-free
and not traced by bright CO emission.
There is growing observational evidence for such ``dark molecular
gas'' \citep{Wolfire2010}.  Comparison of the total hydrogen column
density using gamma-ray \citep{Grenier2005,Abdo2010} or dust emission
\citep[e.g.,][]{Planck2011} with maps of 21 cm \HI\ and CO-bright H$_{2}$
show that there is significant dark H$_{2}$ surrounding most nearby
molecular clouds.  The mass of such CO-free dark gas might be comparable
to the mass of H$_{2}$ traced by CO emission.  The GOT~C$^+$ survey of
the Milky Way disk in the CII 158~$\mu$m line also provides evidence
for a substantial amount of CO-dark H$_{2}$
\citep{Pineda2013}.  Additionally, \citet{Allen2012}
obtained deep 18~cm observations of OH emission in a portion of the
Perseus arm, finding abundant OH in regions that do not show
detectable CO emission.




H$_{2}$ can reach a high equilibrium abundance in the ISM in gas flows
that have lower volume and surface densities than those typical of
GMCs \citep[e.g.,][]{Dobbs2008}.  
For instance, \citet{Clark2012} demonstrate that the formation
of CO-bright regions in the gas occurs only around 2~My before the
onset of star formation. In contrast, H$_2$-dominated but CO-free
regions form much earlier.
The time scales might be long enough that \HI\ and CO-dark gas can coexist,
as evidenced by the massive \HI\ envelopes surrounding the Perseus
\citep{Lee2012} and Taurus molecular clouds \citep{Heiner2013}. The
level of dissociating UV flux is not sufficient to explain all of this
\HI\ as a photo-dissociation product, suggesting that these \HI\
envelopes might be converting into CO-dark gas.




\subsubsection{Conditioning the ISM Reservoir}
\label{sec:assembly}


Large-scale gravitational instabilities in the Galactic disk can drive
GMC formation.
%
Stability of a disk can be quantified in terms of the Toomre parameter
\citep{Toomre1964}
$Q \equiv c_{\rm s} \kappa/\pi G \Sigma$, where $c_{\rm s}$ is the
sound-speed of the gas, $\kappa$ is the epicyclic frequency, which is
typically of the same order of magnitude as $\Omega$, the rotational
frequency of the disk, and $\Sigma$ is the surface density of the
disk. A pure gas disk is gravitationally unstable wherever $Q <
1$. The analysis for a disk that contains both gas and stars is more
complex \citep[e.g.,][]{Elmegreen2011}, but instability
still requires a Toomre parameter of order unity.

The question of whether the Milky Way is Toomre stable has been
examined recently by \citet{Krui+2013}.
They conclude that in the CMZ the gas is marginally unstable, with $Q
\sim 1$, only within the central so-called 100-pc ring (see \S\ref{section_cmz}) revealed by Herschel \citep{Molinari+2011}. 
On larger scales in the disk, they
find stability for $R_{gal} < 4 \: {\rm kpc}$.
\citet{Krui+2013} also find
that the disk is highly stable at $R_{gal} > 20 \: {\rm kpc}$, due to
its low surface density, and although metallicity dependent variations
in the CO-to-H$_{2}$ conversion factor could lead to this being
underestimated \citep{Glover2011}, it is unlikely
to make enough of a difference to render the disk unstable.

It is therefore plausible that on the largest scales, gravitational
instability helps to gather together the gas required for GMC
formation. However, GMCs probably do not form directly from this
instability: the growth time scale is long \citep{Elmegreen2011}, and
the applicability of the Toomre analysis, which assumes an infinitely
thin disk, to scales smaller than the disk scale height is also
questionable. It is far more plausible that gravitational instability
promotes GMC formation indirectly, by enhancing non-axisymmetric
features such as spiral arms that help to gather the diffuse gas
together \citep[e.g.,][]{Dobbs2012}, and by directly driving
turbulence in this gas
\citep[e.g.,][]{Wada2002}. 
Both mechanisms can induce the converging flows that constitute the
last stage of compression necessary to form GMCs.

Another form of mass assembly is bubbles created by massive star
feedback. Giant \HII\ regions and the collective effects of multiple
supernovae in OB associations can sweep up kpc-scale shells which blow
out of the disk, and massive, slowly expanding rings of gas in the
Galactic plane.  The 21~cm surveys of \citet{HartmannBurton1997}
reveal hundreds of \HI\ supershells in the Galaxy surrounding OB
associations \citep{Koo1992,Ehlerova2013}.  
These 0.1 to 1~kpc scale features surround bubbles of hot plasma
traced by extremely low density \HII\ regions and X-ray emission.
They appear to be powered by the combined impact of photoionization,
stellar winds, and supernova explosions. 
\cite{McCrayKafatos1987} 
proposed that gravitational instabilities in the supershells swept-up
by expanding superbubbles would first occur on the mass-scale of GMCs
in the Solar vicinity.  As superbubbles sweep-up the surrounding ISM,
the resulting shell expansion velocities decrease.  
Once the local velocity dispersion of a given patch of the shell drops
below the gravitational escape velocity, that region becomes subject to
gravitational instability.  
Both small and large length- and mass-scales are stable, giving rise
to a critical intermediate scale, which in the Solar vicinity is
similar to the observed scales of GMCs.

Regardless of whether it is ultimately gravity or stellar feedback
that is more important for assembling diffuse gas into dense clouds,
the behavior of the gas on small scales is likely to be rather
similar.  Because GMC formation requires large amounts of gas to be
gathered into one place, GMCs will typically form at the stagnation
points of converging flows of gas, and the details of their formation
will have little dependence on how these flows are driven.

\subsubsection{Large and Small-Scale Converging Flows}
\label{sec:flows}

The potential importance of converging
or ``colliding'' flows as a GMC formation mechanism was first stressed
by \citet{Ballesteros-Paredes+99} and \citet{Hartmann+01}.
The earliest one-dimensional models
\citep[e.g.,][]{Hennebelle+Perault99,Koyama+Inutsuka00}
showed that the collision of streams of thermally stable warm gas
could increase their density sufficiently to render the gas thermally
unstable \citep{Field65}. The resulting thermal instability leads to a
large increase in the gas density from a value $\sim 1 \: {\rm
  cm^{-3}}$, characteristic of the warm neutral medium (WNM), to $\sim
100 \: {\rm cm^{-3}}$, characteristic of the cold neutral medium
(CNM). At the same time, the temperature of the gas falls by two
orders of magnitude, from $T \sim 6000$~K in the WNM to $T \sim 60$~K
in the CNM.

Subsequent 2D and 3D modeling demonstrated that the
thermal instability occurring in the post-shock gas layer naturally
produces turbulence with a velocity dispersion similar to the sound
speed in the unshocked gas \citep[e.g.,][]{Koyama+Inutsuka02, Heitsch+05,
Vazquez-Semadeni+06, Banerjee2009}. 
Thermal instability can create dense structures directly via the
isobaric form of the instability.  However, most of
the resulting small clumps are not self-gravitating
\citep{Koyama+Inutsuka02, Heitsch+05, Glover2007},
and instead it is the ensemble of clumps produced by a coherent
converging flow that might become unstable
\citep{Vazquez-Semadeni+07}.  
%

An important feature of the clouds formed in colliding flows is that
they typically undergo global gravitational collapse primarily in the
directions perpendicular to the flow
\citep[see, e.g.,][]{Vazquez-Semadeni+07}. 
Such collapse naturally leads to the formation of filaments.  The
non-linear development of the thermal instability also proceeds by
forming a network of filaments (see \S\ref{filaments}), at whose intersections clumps (see \S\ref{par_clumps}) are
subsequently formed \citep{Audit2005, Vazquez-Semadeni+06,
  Heitsch+08}.


\subsection{Lessons from the Solar Vicinity}

Currently, the Sun appears to be located in an inter-arm spur of gas
and dust between two major spiral arms of the Galaxy \citep{Xu2013}.
The Sun appears to be inside a superbubble.
Most of the ISM in the Solar vicinity is expanding with a mean
velocity of 2--5~km~s$^{-1}$, discovered in \Hi\ and more
apparent in molecular gas \citep[e.g.,][]{Lindblad1973,dame2001}.  At
the center of ``Lindblad's ring'' is the 50 My old Cas--Tau group, a
``fossil'' OB association \citep{Blaauw1991}, whose most massive 
members have evolved and become supernovae.
%
%
The kpc-scale superbubble produced by the Cas--Tau group blew out of
the ISM both above and below the Galactic plane.  Consistent with
ballistic trajectories of dense clouds entrained in or formed from supershells,
\hi\ surveys reveal complexes of in-falling gas
\citep{Stark1992,Kuntz1996}.

The nearest OB associations such as Scorpius-Centaurus, Per OB2, Orion
OB1, and Lac OB1b, and the B and A stars that trace the so-called
``Gould Belt'' of nearby young and intermediate age stars, are all
associated with Lindblad's ring \citep{Lesh1968,DeZeeuw1999}, evidence
for secondary star formation in clouds that condensed from the
ancient, expanding, and tidally-sheared supershell (see Figs.~5 and 6
in \citealt{PerrotGrenier2003}).  These nearby OB associations have
spawned their own superbubbles with sizes ranging from 100 to 300
pc. The remaining GMCs in each of these regions are embedded within
their respective superbubbles.  Thus most of the star-forming
molecular clouds in the Solar vicinity have been extensively impacted
and processed by massive star feedback from nearby OB associations.

\begin{figure}[h]
\includegraphics[width=0.485\textwidth]{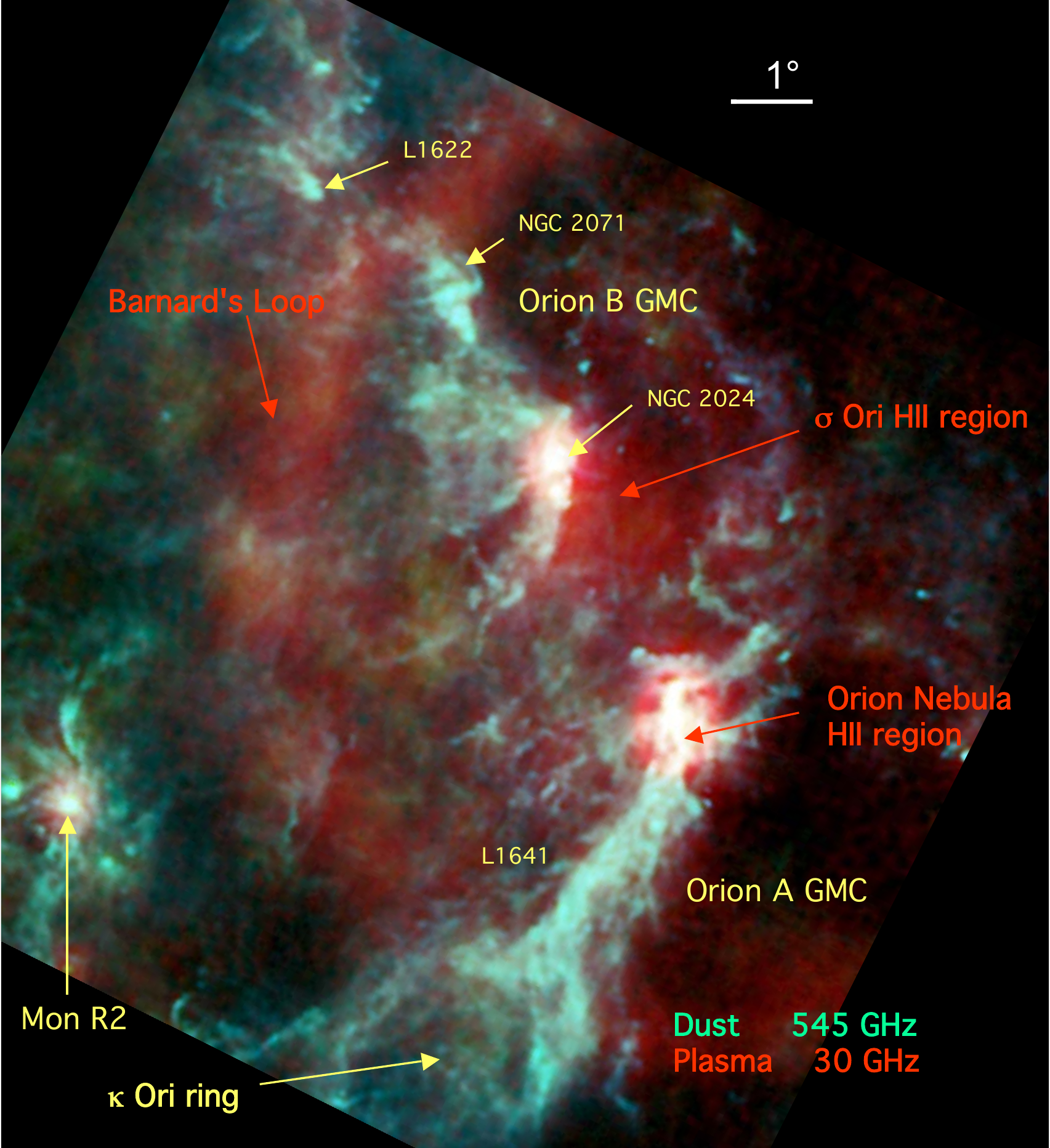}
\caption{Planck view of the $10^5$ \msol\ Orion A and B GMCs,
  traced by dust continuum emission at 545 GHz (turquoise), and the
  eastern portion of the Orion-Eridanus superbubble, in free-free
  emission at 30 GHz (red). The cometary Orion A cloud faces the
  center of the Orion-Eridanus bubble.  The Orion Nebula and the
  OMC1/2/3 clumps are the most active sites of star formation in the
  Solar vicinity.}
\label{fig:Orion_Planck}
\end{figure}

Figure~\ref{fig:Orion_Planck}, after \citet{Bally2010Nature} and
updated using Planck data \citep{planck2013}, shows the
Eastern part of the Orion-Eridanus superbubble in detail.  The
northern portion of the Orion A cloud hosts the most active site of
ongoing star formation within 500~pc of the Sun.  The Orion A cloud is
cometary with its northern end located toward the projected interior
of the superbubble.  The Orion Nebula and the OMC1 core, the closest
site of active high-mass star formation, is in the center of the
degree-long, high-density Integral Shaped Filament (see \S\ref{fil_sub})
\citep{JohnstoneBally1999,JohnstoneBally2006}.

\subsection{Lessons from the Central Molecular Zone}
\label{section_cmz}

\begin{figure*}[ht]
\includegraphics[width=1.0\textwidth]{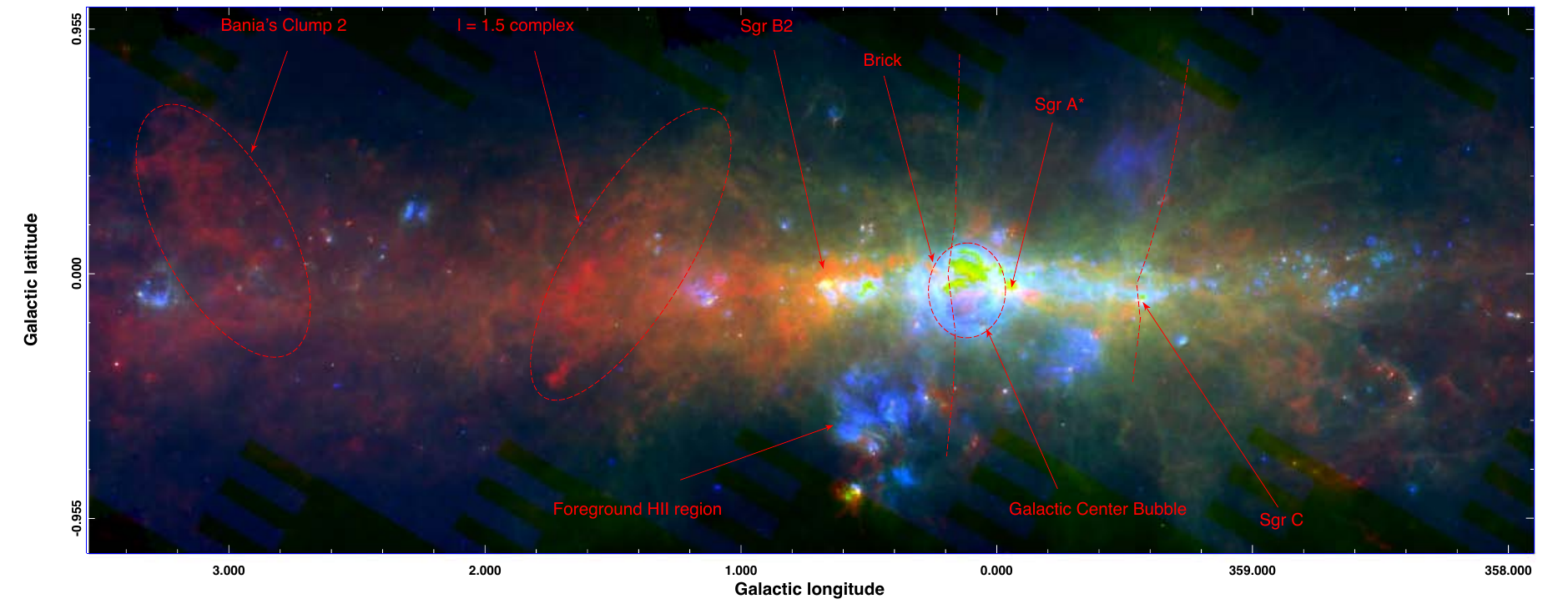}
\caption{The Central Molecular Zone (CMZ) in a composite three-color
  image combining {\it Spitzer} 24 $\mu$m (blue) with {\it Herschel} 70
  $\mu$m (green) and 500 $\mu$m (red). Noticeable features are identified in the figure; the Sofue-Handa lobe structure (see text) is delimited by the two nearly vertical red-dashed lines at $l\sim$0.2\deg, and $l\sim -0.5$\deg.
 }
\label{fig:GC_wide}
\end{figure*}





Extending from longitudes $\sim -5^\circ$ to $+5^\circ$
(Fig.~\ref{fig:GC_wide}), the CMZ provides a unique opportunity to
investigate the properties of dense gas in a circumnuclear environment
containing a supermassive black hole (SMBH).  Approximately 10\% of
the Milky Way's ISM lies within 0.5 kpc of the Galactic Center's SMBH.
The CMZ contains the most extreme GMCs and star forming regions in the
Galaxy.
%

The CMZ contains  2 - 6 $\times 10^7$ M$_{\odot}$ of molecular gas 
with average densities greater than $10^4~ \rm cm^{-3}$, sufficient 
to excite most high dipole-moment molecules such as NH$_3$, 
CS, HCN, HCO$^+$, and HNCO everywhere in the CMZ 
\citep{Dahmen1998,Ferriere2007,Jones2012}. The CMZ 
GMCs differ from GMCs
in the Galactic disk in several ways:  1) They have one to two 
orders of magnitude higher mean density. 2) High dipole-moment 
molecules are excited throughout the volume of CMZ GMCs, not 
just in isolated clumps and cores as in Galactic disk GMCs. 3) CMZ 
GMCs have larger line-widths (typically 15 to more than 50 km s$^{-1}$) 
than disk GMCs (typically 2 to 5 km s$^{-1}$). Large line-widths are 
an indication of large-amplitude internal motions. Indeed, 
\citet{Huettemeister1998} found extensive SiO emission from the 
CMZ clouds, indicating the presence of shocks sufficiently strong to 
sputter dust grains to produce a significant gas-phase abundance of SiO.
Molecular gas in the CMZ is not only dense, but also warm, and able to
excite high-J transitions of CO, as well as a variety of high dipole-moment molecules
\citep{Martin2004}.  

\begin{figure}[h]
\includegraphics[width=0.49\textwidth]{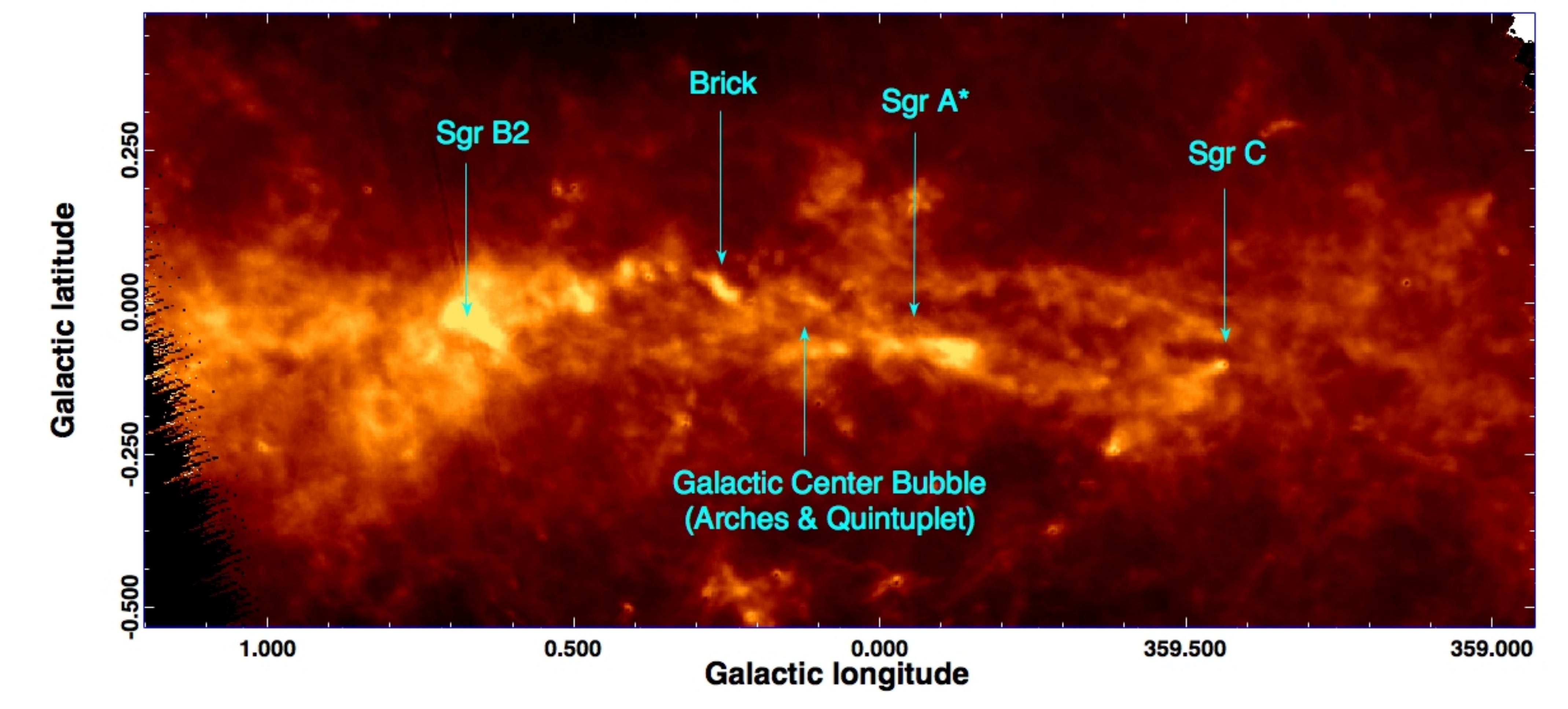} \\ 
\includegraphics[width=0.49\textwidth]{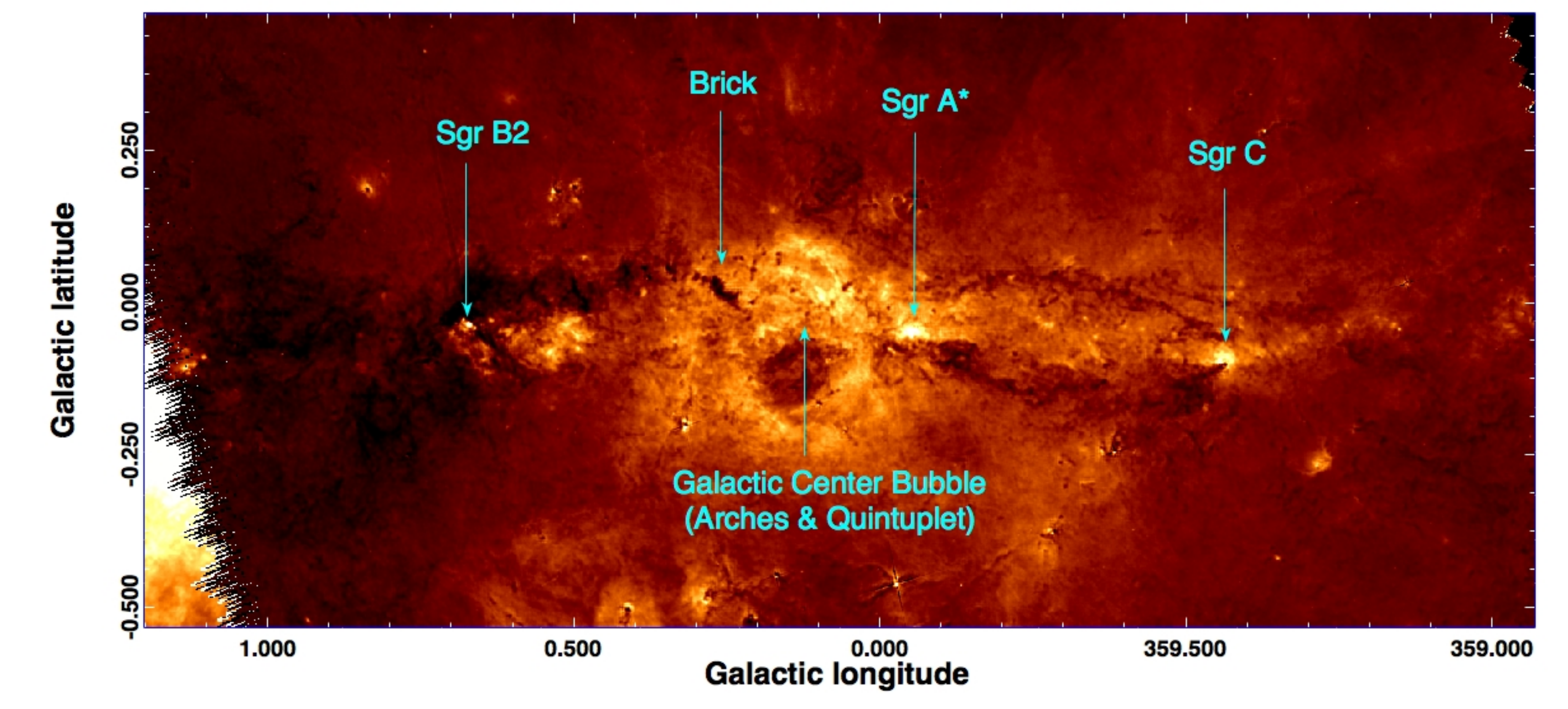}
\caption{Maps of background-subtracted dust column density (upper) and
  dust temperature (lower) from Hi-GAL images \citep{Molinari+2011} of
  the inner 2\fdg2 of the CMZ.  Large regions of cold dust are
  associated with Sgr B2 (left), Sgr C (right), and the 20 km~s$^{-1}$
  cloud (right of center).  The warm Galactic Center Bubble is to the
  left of Sgr A.}
\label{fig:GC_Herschel}
\end{figure}

The orbital dynamics of the gas is critical to the evolution.
Non-self-intersecting orbits form two families: the x1 orbits whose
major axes are aligned with the bar and the x2 orbits which tend to be
elongated orthogonal to the bar.  The x1 orbits become more ``cuspy''
as their semi-major axes shrink, eventually becoming
self-intersecting.  Clouds on such orbits will collide, resulting in
loss of orbital angular momentum and settling onto the more
compact x2 family of orbits.  While the x1 orbits are populated by \HI\
and some molecular clouds, the x2 orbits contain the densest and most
turbulent molecular clouds in the Galaxy.  Almost all star formation
in this gas, heating the warm dust seen as bright blue and green
emission in Figure~\ref{fig:GC_wide}, occurs within about 100 pc of
the nucleus.
The CMZ gas is asymmetrically located with respect to the dynamic
center of the Galaxy.  Two-thirds of this emission is at positive
longitudes and has positive radial velocities; only one-third is at
negative longitudes and has negative radial velocities
\citep{Bally1987b, Bally2010_CMZ,Jones2012}. Remarkably, the majority
of 24 $\mu$m sources in Figure~\ref{fig:GC_wide} are located at {\it
  negative} Galactic longitudes, not on the side of the nucleus
containing the bulk of the dense CMZ gas and associated cold dust
(red).
\cite{Molinari+2011} found that the coolest dust and the highest
column density gas is located in an {\it infinity-sign-shaped} ring
with a radius of about 100 pc (Fig.~\ref{fig:GC_Herschel}).

Other notable features are marked in these figures.  The Sgr B2
molecular cloud at ${\it l}$ = 0.8$^\circ$ contains over $10^6$
M$_{\odot}$ of molecular gas and hosts what might be the most luminous
site of massive star formation in the Galaxy.  It is located adjacent
to the older Sgr B1 region at the high longitude end of the 100 pc
ring. Sgr C, located at the negative longitude end of the ring, hosts
the second most massive site of star formation in the CMZ.

The prominent oval of warm dust (left of center in the lower panel of 
Fig.~\ref{fig:GC_Herschel}) is a very young 30 pc diameter
superbubble (the ``Galactic Center Bubble" or GCB) whose interior
contains the massive, 3 to 5 My old Arches and Quintuplet clusters.
The Sgr A region and the SMBH are located outside the GCB; thus the GCB
might be powered mostly by massive stars.  The GCB is the smallest
and brightest member of a set of nested bubbles emerging from the
central 100 pc region of our Galaxy.  The Sofue-Handa lobe
\citep{SofueHanda1984} is a degree-scale bubble traced by free-free
emission blowing out of this region.  The relationship between
these features and the recently recognized kpc-scale Fermi-LAT-Planck
bubble \citep{Su2010,planckhaze2013} remains unclear.  Is this bubble
powered by the merging of superbubbles fed by dying massive stars in
the CMZ, fed by occasional AGN activity of the SMBH, or a combination
of these mechanisms?


The compact, high column density clump of cold dust located to the
left of the GCB (``Brick'' in Figs.~\ref{fig:GC_wide} and \ref{fig:GC_Herschel}) is the most
extreme IRDC in the Galaxy (Longmore et al., this volume).  This clump
is located in the 100 pc ring, has a mass greater than $10^5$
M$_{\odot}$ within a radius of 3 pc, yet shows no evidence of any star
formation \citep{2012ApJ...746..117L}. It is part of a string of
massive clouds stretching from a location near Sgr~A$^*$ to Sgr~B2,
including several with various degrees of on-going star formation,
possibly triggered by a close passage near the SMBH
\citep{2013MNRAS.433L..15L}.
 

Wide-field mapping of the CMZ in the dust continuum, high-J CO lines,
the CI line, and a variety of dense gas tracers will shed light on a
number of critical questions about star formation, the ISM, and its
interactions with the SMBH.  Does nuclear star formation shield the
central black hole from growth?  Or does occasional flaring of the
SMBH regulate the state of the ISM and star formation in the CMZ? What
is the star formation rate in the CMZ?  Does it follow the
Schmidt-Kennicutt relation, or is the star formation rate per unit
mass of dense gas lower in the CMZ than in galactic disks
\citep{Longmore2013, Krui+2013}?







%
%
%



\section{Converting GMCs  to Dense Structures}
\label{dense_structures}

It is well established that only a small fraction of the gas in GMCs
is dense enough to be involved in star formation.
This is apparent from CS surveys in high-mass star forming regions
\citep[e.g.,][]{Plume1992,Plume1997}, as well as from large-scale gas
and dust mapping of nearby low-mass star forming regions (e.g.,
\citealt{Goldsmith2008}).

Herschel surveys allow investigation of the properties of IRDCs
because the cold and dense dust shines brightly in the submm
continuum.  Visual inspection by \cite{Wilcock+2012} of the Hi-GAL
survey in the 300\deg $\leq l \leq$ 330\deg\ region of the Galactic
Plane suggests that only 50\% of the \citet{Peretto+Fuller2009} IRDCs
are \textit{bona fide} cold and dense structures while the rest are
regions devoid of background mid-IR emission.  Nevertheless, the
Peretto \& Fuller column densities derived from the mid-IR opacity
provide independent measurements that can be compared with estimates
obtained from Herschel's complete far-IR photometric coverage
\citep{Elia+2013, Elia+2014b, Schisano+2014}.

Unbiased surveys allow statistically significant estimates of the
lifetimes and relative durations of the various stages of evolution of
clouds, clumps, and cores.  For example, \citet{Tackenberg+2012} use a
portion of the ATLASGAL survey to estimate the time scale for massive
starless clumps using association to GLIMPSE sources, obtaining quite
short (6--$8 \times 10^4$~y) time scales. \cite{Wilcock+2012}
analyse a portion of the Hi-GAL survey and find that only 18\% of the
Peretto \& Fuller \textit{bona fide} IRDCs in the 300\deg $\leq l
\leq$ 330\deg\ range appear not to be associated with either 8 or
24\mum\ counterparts.

An interesting approach has been adopted by \cite{Battersby+2011}, in
which Hi-GAL column density and temperature maps in the $l$=30\deg\
and $l$=59\deg\ fields were compared to companion maps in which each
pixel is populated with a variety of star formation indicators,
including 8 and 24 \mum\ emission, association with methanol masers and
Extended Green Objects suggestive of the presence of outflows. A
pixel-to-pixel comparison showed that pixels with more star formation
indicators have higher column densities and higher temperatures with
$T_{dust} >$ 20~K.

\begin{figure*}[t]
\includegraphics[width=1\textwidth]{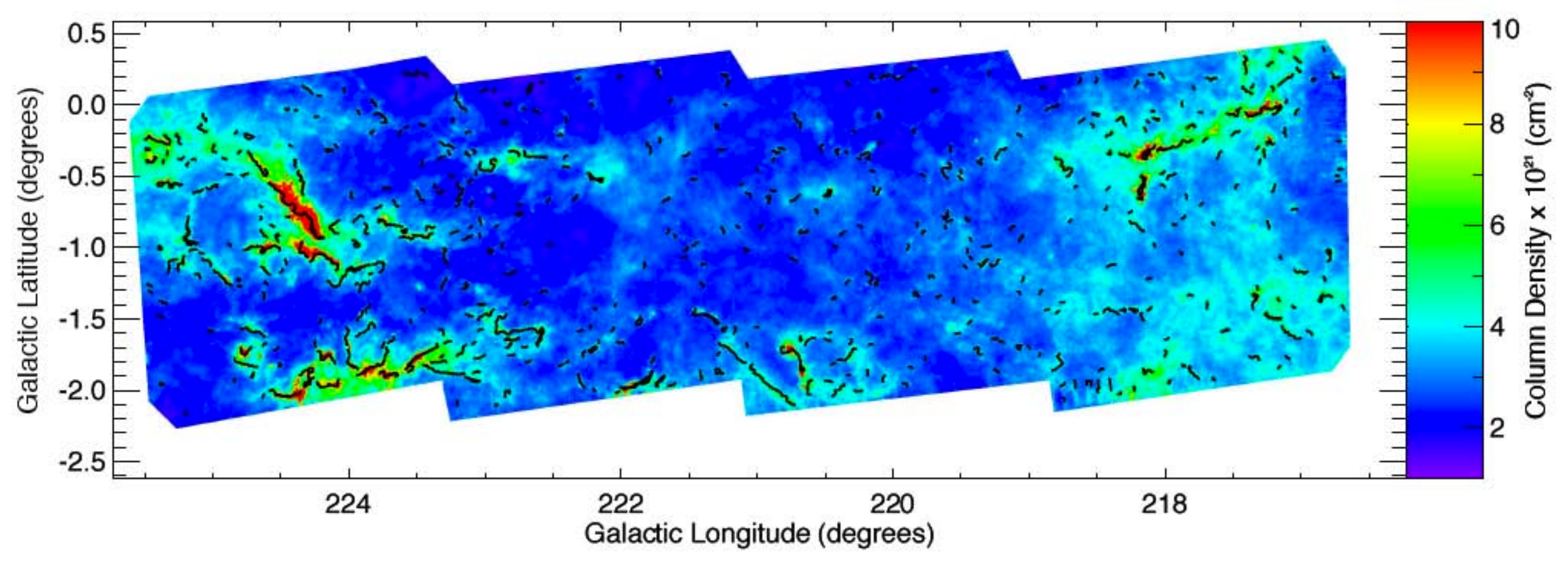}
\caption{Column density map from Hi-GAL images in the Galactic
longitude range 217\deg$\leq l \leq 224$\deg \citep{Elia+2013}.  The
thick black lines delineate the main spine of each detected filament
(branches are neglected for clarity). Detected filaments are those
that show a contrast greater than three times the standard deviation of the column density 
computed locally.}
\label{fig:eu_fig10}
\end{figure*}

\subsection{Filaments}
\label{filaments}

Surveys with Herschel reveal that the densest regions in molecular
clouds are organized in complex filamentary structures
\citep{Molinari2010b}.  As discussed above and by Andr\'e et al.\
(this volume), nearby clouds such as Taurus and Ophiuchus are
filamentary and Hi-GAL reveals similar structures on much larger
scales (several to tens of parsecs) along the Galactic Plane.  The ISM
is permeated by a web of filamentary structures from very low column
density ``cirrus" emission to gravitationally ``supercritical" clouds,
and from the small scales of star-forming cores to scales of tens to
hundreds of parsecs in the Galactic Plane.

PACS and SPIRE maps at 70 - 500\mum\ showed that the Aquila Rift and
the Polaris Flare contain an extensive network of filaments, many of
which are several pc in length \citep{andre10}.  In IC 5146
\citet{Arzoumanian2011} show that the filaments have radial density
profiles that fall off as $r^{-1.5}$ to $r^{-2.5}$ and mean FWHM
widths of $0.10 \pm 0.03$~pc.  This is much narrower than the
distribution of Jeans lengths which range from $\sim 0.02$--0.3~pc.
Because the filament widths are within a factor of two of the sonic
scale at which the turbulent velocity dispersion equals the sound
speed, they suggest that large scale turbulence could be responsible for
forming the filaments whereas gravity then drives their evolution by
accreting material from the surroundings.  This idea seems to be at
least circumstantially supported by \cite{Chapman2011} whose
polarization map of the L1495/B213 filament shows a number of
sub-filaments oriented perpendicular to the main filament, but aligned
with the magnetic field direction as measured in the surrounding
material.  These sub-filaments might be related to mass accretion onto
the main filament \citep{Palmeirim2013} which, in turn, might help to
form the dense cores within \citep{Hacar2011}.

\citet{Fischera2012, Fischera2012b} discuss equilibrium models of
filaments as isothermal self-gravitating infinite cylinders in
pressure equilibrium with the ambient medium.  These offer an
analytical and quantitative explanation of the radial profiles and the
width--column-density relation.  However, there is a maximum mass line
density $M_{\rm lin, max}$ beyond which there is no equilibrium
solution (often referred to as ``critical'').  If the filaments have
only thermal support, then at 10~K $M_{\rm lin,max} \sim 16 M_\odot$
pc$^{-1}$, corresponding to a maximum column density of $N_{\rm
  H_2,max} \sim 10^{22}$ cm$^{-2}$ or $A_{\rm V,max} \sim 10$.  Note
that if there were in addition \emph{turbulent} support, as seems
relevant to accreting filaments, then a quasi-static state might exist
with higher $M_{\rm lin}$.  Filaments close to, or exceeding, $M_{\rm
  lin, max}$ are subject to fragmentation into self-gravitating
prestellar cores \citep{Fischera2012}.
Given the derived \emph{dust} temperature in these filaments, and
assuming that the gas temperature is similar, \citet{andre10} find
that 60\% of the bound (self-gravitating) prestellar cores in these
regions are located in filaments with $M_{\rm lin} > M_{\rm lin,
  max}$, indicating a strong association of supercritical filaments
and star formation. On the other hand, while filaments with $M_{\rm
  lin} < M_{\rm lin,max}$ interestingly contain cores, few of these
cores have strong self-gravity.

Herschel has shown that filamentary structures are organized into
complex interconnected branches and also loops.  Examples include the
Vela region \citep{Hill+2011} and DR21 \citep{Schneider+2010}.  The
first results from the Hi-GAL survey showed widespread filamentary
structures on larger scales in a $2^\circ \times 2^\circ$ region
toward $l = 59^\circ$ \citep{Molinari2010b}.  Like \cite{andre10},
they find that dense cores/clumps tend to be associated with the
filaments. They find that the column densities of the filaments lie in
the range $10^{21}$ cm$^{-2} \le$ N(H$_2$) $\le 10^{22}$ cm$^{-2}$,
corresponding to $1 \le A_V \le 10$. This range is lower than found
for local clouds like the Aquila Rift (distance 260~pc) and the
Polaris Flare (150~pc) and could be due in part to the greater distances
of the filaments in the $l = 59^\circ$ region ($\sim 2$--8~kpc) and
the resulting beam dilution (at these distances, Hi-GAL also detects
clumps instead of cores).

More rigorous detection and quantitative assessment of filaments is
now underway using a variety of tools based on a variety of
morphological analyses of multi-scale diffuse emission.  One of these
approaches is based on the analysis of the Hessian matrix computed
over large-scale column density maps (but see also
\citealt{Arzoumanian2011}), and it has been applied by
\cite{Schisano+2014} over an 8\deg\ x 2\deg\ Hi-GAL column density map
of the Galactic Plane over the longitude range $l = 217^\circ -
224^\circ$.  They identified over 500 filaments (with
$\sim$ 2000 branches) with lengths from 1.5--9~pc and column densities
from $10^{19}$ to $10^{22}$ cm$^{-2}$ (see 
Fig.~\ref{fig:eu_fig10}). Again, these column densities are lower
than those cited by \cite{andre10} for two possible reasons: beam
dilution caused by the greater distances to these filaments (500~pc --
3.5~kpc) and the fact that \cite{Schisano+2014} use the average column
density rather than the column density at the spine
of the filaments. This latter effect tends to decrease the quoted
column density by a factor of four.  Focusing on only the nearest
filaments, they find average lengths, widths, and column densities to
be 2.6 pc, 0.26 pc, and $9.5 \times 10^{20}$ cm$^{-2}$, respectively.
Nevertheless, these filaments are still wider and less dense than
those in \cite{Arzoumanian2011} by a factor of 2--3, and so not all
demonstrably supercritical.

The analysis of \citet{Schisano+2014} also deals with the physical
properties of what they call the individual ``branches" of the
filamentary networks, i.e., the sub-portions of filamentary structures
limited by the nodal points created by the intersections. They found
that the branches containing protostellar clumps have $M_{\rm lin}
\sim 20 - 30 {\rm ~M}_\odot$ pc$^{-1}$ (arguably supercritical),
whereas branches with only prestellar clumps have $M_{\rm lin} \sim
4 \, {\rm M}_\odot$ pc$^{-1}$.  Although ``prestellar" clumps are
simply those with no counterpart in the Hi-GAL 70 \mum\  band, the
absence of associated 70 \mum\ emission is an indication of
considerably lower rates of star formation than found in protostellar
clumps containing detectable 70 \mum\ sources.
 
 \subsubsection{Filament Formation Mechanisms}

There are a number of physical mechanisms that can form filamentary
structure.  These include gravity, compressive flows in supersonic
turbulence and colliding sheets, production of tails by shadowing by
dense clumps (the pillars of M16 are a well-known example),
fragmentation of expanding shells and supershells, generation of
fingers through Rayleigh-Taylor instability, and magnetic fields.

{\it Gravity:} \citet{Lin1965} demonstrated that spheroidal gas clouds
tend to flatten as they collapse, yielding sheet-like structures, and
sheets collapse into filaments \citep[see
also][]{Larson1985}. 
To produce sheets and filaments, the mass of the collapsing cloud must
be much larger than the Jeans mass, because in clouds with masses
close to $M_{\rm J}$, isotropic pressure forces tend to keep the
clouds quasi-spherical.

{\it Supersonic turbulence:} Both purely hydrodynamic (HD) and
magnetohydrodynamic (MHD) supersonic turbulence can lead to filament
formation \citep[e.g.,][]{Padoan+2001}.  Supersonic turbulence creates
a complex network of shocked-layers confined by the ram pressure of
the convergent parts of the flow.  The supersonic collision between
shock-formed sheets produces denser filaments.  If these
shock-compressed layers become sufficiently massive, their
self-gravity can
lead them to collapse. Otherwise, they will tend to dissolve on a
turbulent crossing time.
The Mach number of the turbulence determines the density contrast
$\rho_2/\rho_1$ between pre-and post-shocked gas: for HD shocks,
this scales as ${\cal M}^{2}$, where the Mach number ${\cal M} = V /
c_s$. For MHD shocks, $\rho_2/\rho_1 \approx {\cal M}$
\citep{Padoan+Nordlund02}, with ${\cal M}$ being given by $V / c_A$ in
this case.
Here, 
$c_s$ is the sound speed and $c_A = B / (4 \pi \rho)^{1/2}$ is the
Alfv\'en speed.  Highly supersonic turbulence therefore yields high
density contrasts, while transonic or subsonic turbulence creates only
small perturbations that then must be amplified by some other
mechanism, such as gravity.  The mass accumulated in post-shock layers
depends on the correlation length of the turbulent velocity field;
larger correlation lengths imply higher mass accumulation layers.

Both gravity and supersonic turbulence formation mechanisms work more
readily in cold and dense gas with a low sound (or Alfv\'en) speed and
small Jeans mass.  Therefore the thermal evolution of the gas plays an
important role.  Rapid and efficient cooling can lead to non-linear
development of thermal instability that can enhance filament formation
\citep[e.g.,][]{Vazquez-Semadeni+00, Vazquez-Semadeni+06,
  Heitsch+05}. Which of these processes
(gravity, turbulence or thermal instability) is primarily responsible
for the formation of the filaments observed in the ISM remains
undetermined; plausibly, all three processes could play important
roles. Nevertheless, recent simulations \citep[e.g.,][]{Gomez_VS13}
suggest that
large-scale gravitational collapse plays the dominant role here.

{\it Shadowing and cometary clouds:} Many \HII\ regions exhibit
cometary clouds with long filamentary tails pointing away from the
direction of ionization.
For example, Figure~\ref{fig:W345} shows several cometary clouds and
filamentary tails in the W4 and W5 giant \HII\ regions.  Figure
\ref{fig:Orion_Planck} shows that the entire Orion A cloud is cometary
with streamers trailing away from the dense Integral Shaped Filament
containing the Orion Nebula Cluster and the OMC1, 2, and 3 cloud cores
which together have spawned over 2,000 YSOs in the last few My.

{\it Fragmentation of expanding shells and supershells:} Bubbles
created by the non-ionizing radiation field of late B and A stars and
\HII\ regions as well as superbubbles and supernova remnants can
sweep-up ambient gas into expanding shells (see \S\ref{sec:assembly}).  Limb brightening of the
shells at the edges can make them look filamentary.  Hi-GAL data
reveals many approximately circular or semi-circular filaments such as
shown in Figure~\ref{fig:NGC7538}.  Additionally, various hydrodynamic
instabilities such as the Vishniac or Rayleigh-Taylor instability, or
the over-running of dense clumps can lead to filamentary structure.
The presence of a regular entrained magnetic field can also confine
the motion of the charged component, thereby breaking the symmetry to
produce filamentary structure.

{\it Rayleigh-Taylor instability:} Rayleigh-Talyor (RT) instabilities
occur when a dense fluid overlies a light fluid, or when there is an
acceleration of a light fluid though a dense layer.  When the shock
fronts associated with expanding giant \HII\ regions or superbubbles
reach radii comparable to or larger than the scale-height of the
Galactic gas layer,
they can accelerate as they encounter an exponentially decreasing
density profile.  Additionally, ionization fronts propagating in a
density profile that decreases faster then $r^{-2}$ tend to become RT
unstable.  The RT instability creates long fingers of dense material
oriented parallel to the pressure or density gradient, with the
appearance of elongated, filamentary, cometary clouds.  Dense clouds
located more than a few scale heights above the Galactic mid-plane
tend to become elongated orthogonal to the plane due to the effect of
superbubbles breaking out of the Galactic gas layer.

{\it Magnetic fields:} In the Taurus clouds, interstellar polarization
of background stars reveals that the minor filaments that run
orthogonal to the dense Taurus filaments \citep{Goldsmith2008} are
parallel to the direction of the field over most of the Taurus
constellation.  This morphology suggests that the charged component of
the gas associated with the minor filaments is magnetically confined.
The charged component is free to stream along the magnetic field lines
and prevented from crossing field lines, instead gyrating about the
field.  Ion-neutral collisions effectively couple the neutral atoms
and molecules to the field.

\subsubsection{Filament Substructure}
\label{fil_sub}

Filaments are observed to have compact substructures consisting of
clumps and cores \citep{Schneider_Elmegreen79, Myers2009, andre10,
  Molinari10}, as expected at least for those that are strongly
self-gravitating (with $M_{line}$ near or
exceeding $M_{line,max}$), but actually present in most.

\begin{figure}[h]
\includegraphics[width=0.485\textwidth]{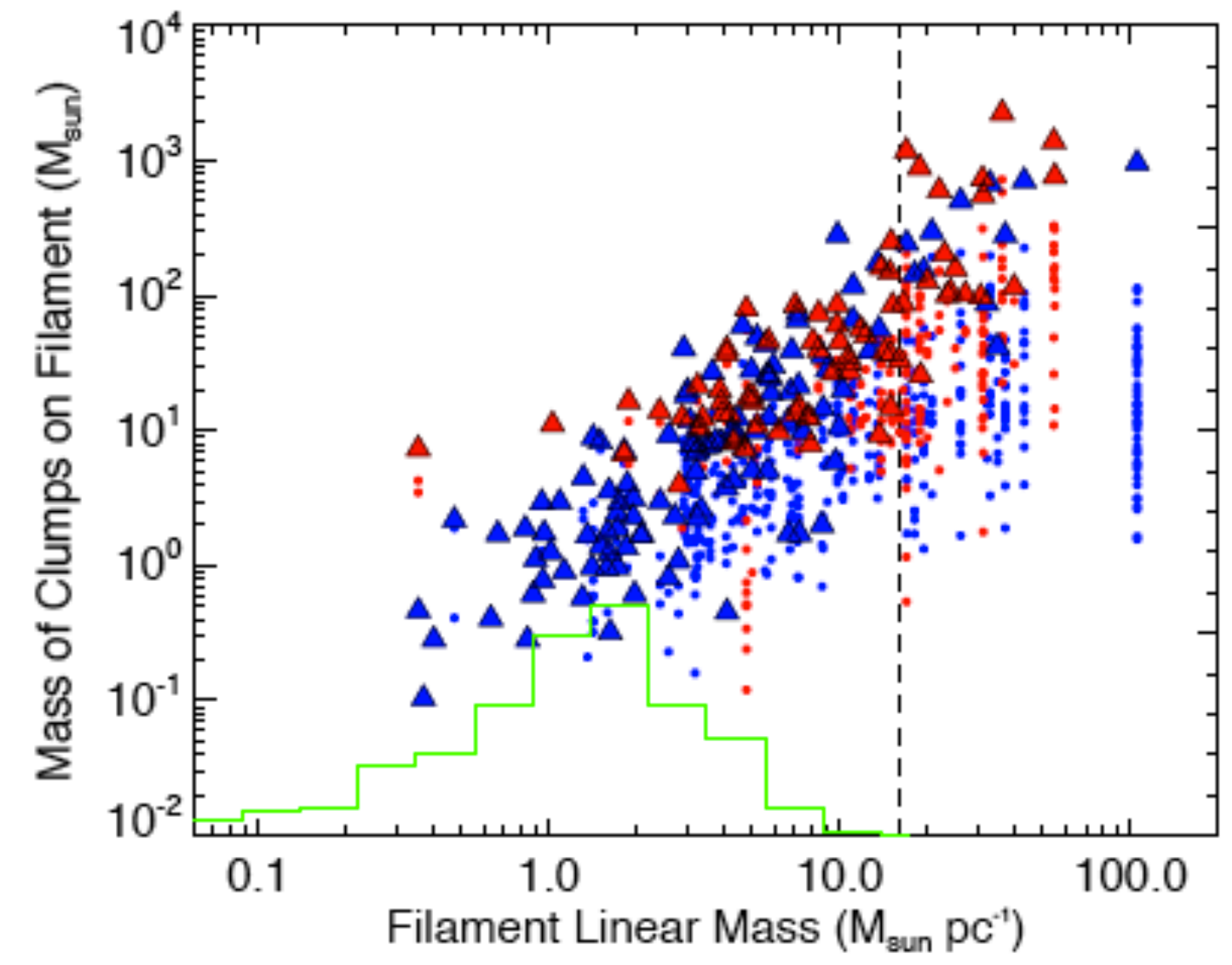}
\caption{Mass of the clumps located in detected filaments in the $l
=217$\deg\ -- 224\deg\ Galactic plane region (\citealt{Schisano+2014},
Fig.~\ref{fig:eu_fig10}) vs.\ $M_{line}$ of the hosting
filament. Small dots are individual clump masses, in red for distance
$d > 1.5$~kpc and blue for $d \leq 1.5$~kpc. Triangles depict the
total mass in clumps on that filament. The distribution of $M_{line}$
for filaments without clumps is also reported in green. The vertical
dashed line marks $M_{line,max} \sim 16$ \msol\ pc$^{-1}$ for $T \sim
10$ K.}
\label{fig:eu_fig18}
\end{figure}


Figure~\ref{fig:eu_fig18} shows a clear relationship between the
masses of the clumps and $M_{line}$ of the filaments hosting
them. $M_{line}$ ranges over more than two orders of magnitude and the
relationship extends well above $M_{line,max}$ \citep{Schisano+2014}.
This correlation might indicate that a spectrum of filament mass line
densities is produced by large-scale turbulence or gravitational
collapse, with clumps almost immediately fragmenting so that their
mass distribution would tend to carry memory of the gravitational
state of the parent filament.

Alternatively,  the correlation 
might result from evolution of filaments and clumps toward larger
masses as accretion continues from the surrounding environment.
Observations suggest continuous flow {\it onto} filaments as well as
from filaments onto cores \citep{Schneider+2010, Kirk+13}.  Numerical
simulations of non-equilibrium collapsing clouds show similar behavior
\citep{Gomez_VS13}.
In such dynamical evolution, filaments fragment into denser clumps
that collapse on shorter time scales than the filaments because of
different dimensionality: collapse time scales for finite filaments
are longer than the clump free-fall time by a factor depending on
their aspect ratio \citep{Toala+12, Pon+12}.  This implies that the
clumps begin forming stars before the filaments have finished
funneling the entire reservoir of gas onto the cores.

The kinematics of the Integral Shaped Filament (ISF) in the Orion A
molecular cloud that contains the Orion Nebula Cluster (ONC;
Fig.~\ref{fig:Orion_Planck}) exhibits this behavior.  The $\sim 7$~ pc
(60\arcmin) long and 0.2 pc wide ISF exhibits a velocity jump of 2 to
3 \kms\ at the position of the OMC1 cloud core located immediately
behind the Orion Nebula.  The southern half of the ISF has a radial
velocity of $V_{LSR} \approx 8$~\kms\ while just north of OMC1
$V_{LSR}$ jumps to about 10~\kms; $V_{LSR}$ increases to over 12~\kms\
near the northern end of the filament adjacent to the NGC1977 \HII\
region \citep{Ikeda2007_OrionA}.  Despite the unknown projection
effects, this velocity jump near the location of the Orion Nebula
appears to be similar to the escape velocity with respect to the mass
of the OMC1 cloud core and the ONC
at a radius of 1 pc from the cluster core, about 3~\kms .  Thus, it
seems likely that material in the ISF is being drawn along its length
by the gravity of the ONC and the adjacent OMC1 core.  If this picture
is correct, then the smaller OMC2 and OMC3 clusters might be dragged
into the ONC, helping to fuel continued growth of the stellar
population there.
The time scale for such a merger is roughly 1~My, 
longer than the crossing time of the ONC ($< 0.3$~My) or the OMC1
core ($< 0.05$ My).

\begin{figure}[h]
\includegraphics[width=0.485\textwidth]{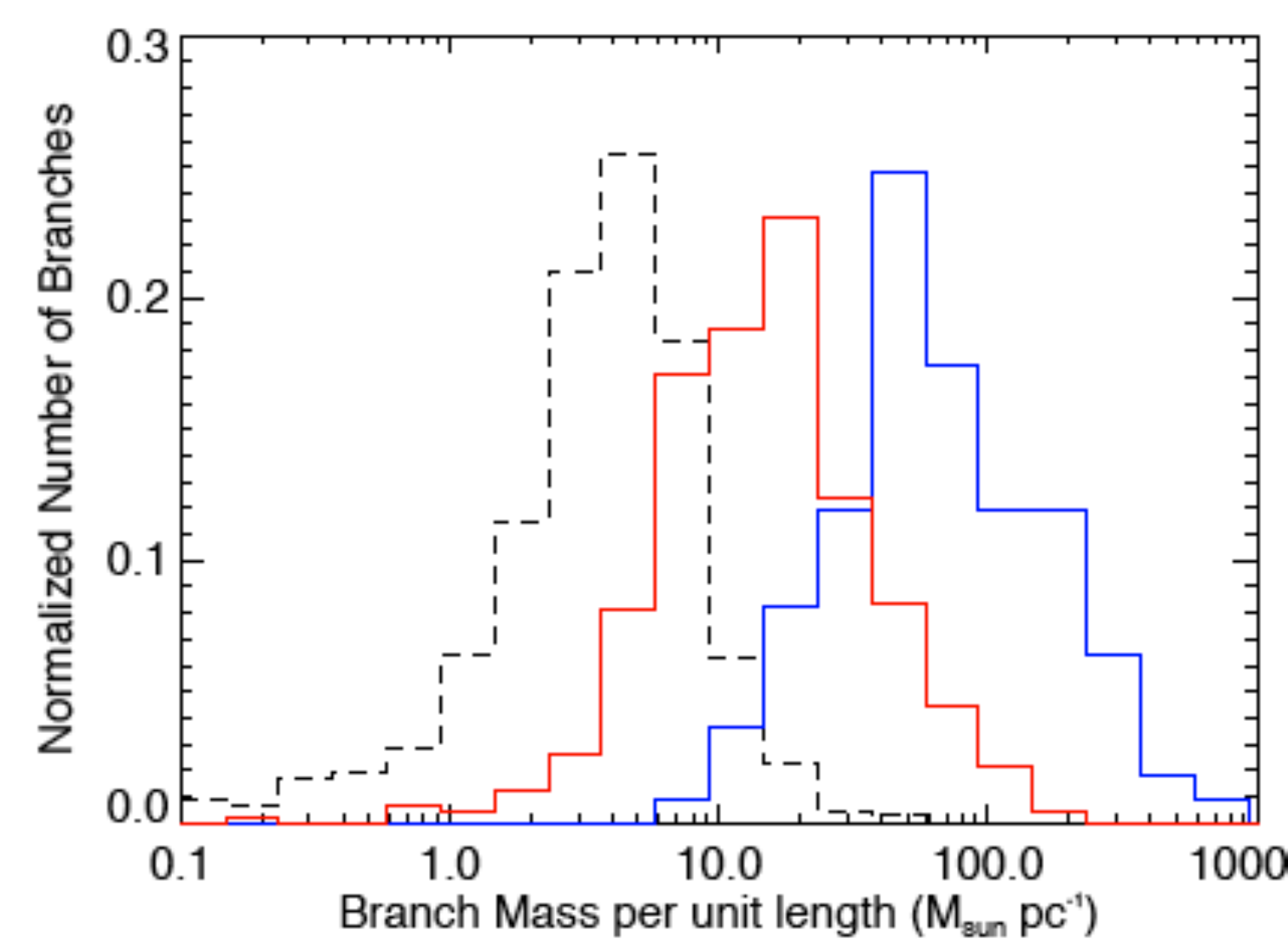}
\caption{Histograms of $M_{line}$ of filament branches detected by
\citet{Schisano+2014}: with protostellar clumps (blue solid line),
with prestellar clumps (red line), and without clumps (black dashed
line). Among the latter, those with higher $M_{line}$ are part of more
complex filaments that host both prestellar and protostellar
clumps. The histograms have been normalized to the respective total number of
filaments to emphasize intrinsic differences between the
distributions.}
\label{fig:eu_fig20}
\end{figure}

The histograms of $M_{line}$ in Figure~\ref{fig:eu_fig20} show
systematic differences depending on the presence of prestellar to
protostellar clumps.  If the filament branches with prestellar
clumps are to evolve into the branches with protostellar clumps, then
they would have to accumulate enough mass for an order of magnitude
increase in $M_{line}$ within only the prestellar core lifetime of a
few times $10^{5}$~years, channeling a fraction of this material onto
the clumps, thus increasing their mass too. This would imply
extremely high accretion rates of the order of $\sim 10^{-3}/10^{-4}$
\msol\ pc$^{-1}$ y$^{-1}$ from the surrounding environment and along
the filament branches.  Orion's ISF has a mass of order $3 - 5 \times
10^3$ \msol.  An accretion time scale of 1 My implies an accretion
rate into the ONC of about 3--$5 \times 10^{-3}$ \msol\
y$^{-1}$. Such rates, although high,
seem consistent with recent observations of the
SDC335.579-0.272 cloud by \cite{Peretto+2013}.  In this scenario,
filaments such as the ISF are long-lived entities that become denser
and more massive over time.

\cite{Schisano+2014} find no correlation between $M_{line}$ of a
branch and the clump formation efficiency within that branch.
However, the $L/M$ ratio of protostellar clumps increases with
$M_{line}$, suggesting that the increased $M_{line}$ in branches
containing protostellar clumps results in an increased star formation
rate.  Contrary to previous proposals, $M_{line}$ might not always
trace the evolution of a filament.  When filaments form from
large-scale turbulence, many might be transient objects, with only
those that are strongly self-gravitating
forming stars.


\subsection{Clumps}
\label{par_clumps}

\subsubsection{Observed properties} 

Clumps and cores have been the subject of star formation studies for
decades (see, e.g., \citealt{Lada+07} or the reviews by Andre et al.\
and Offner et al.\ in this volume).
In nearby clouds, the core mass spectrum closely resembles the stellar
IMF \citep{Motte+1998, Alves+07}, leading to arguments that it
determines the IMF. Such detailed studies are limited to low-mass
objects, less than about 10 \msol. Estimates toward more distant
massive clumps are less definitive due to the relatively limited
spatial resolution available until now for large samples
(\citealt{Shirley+2003, Reid-Wilson2005, Beltran+2006}).

\begin{figure}[h]
\includegraphics[width=0.485\textwidth]{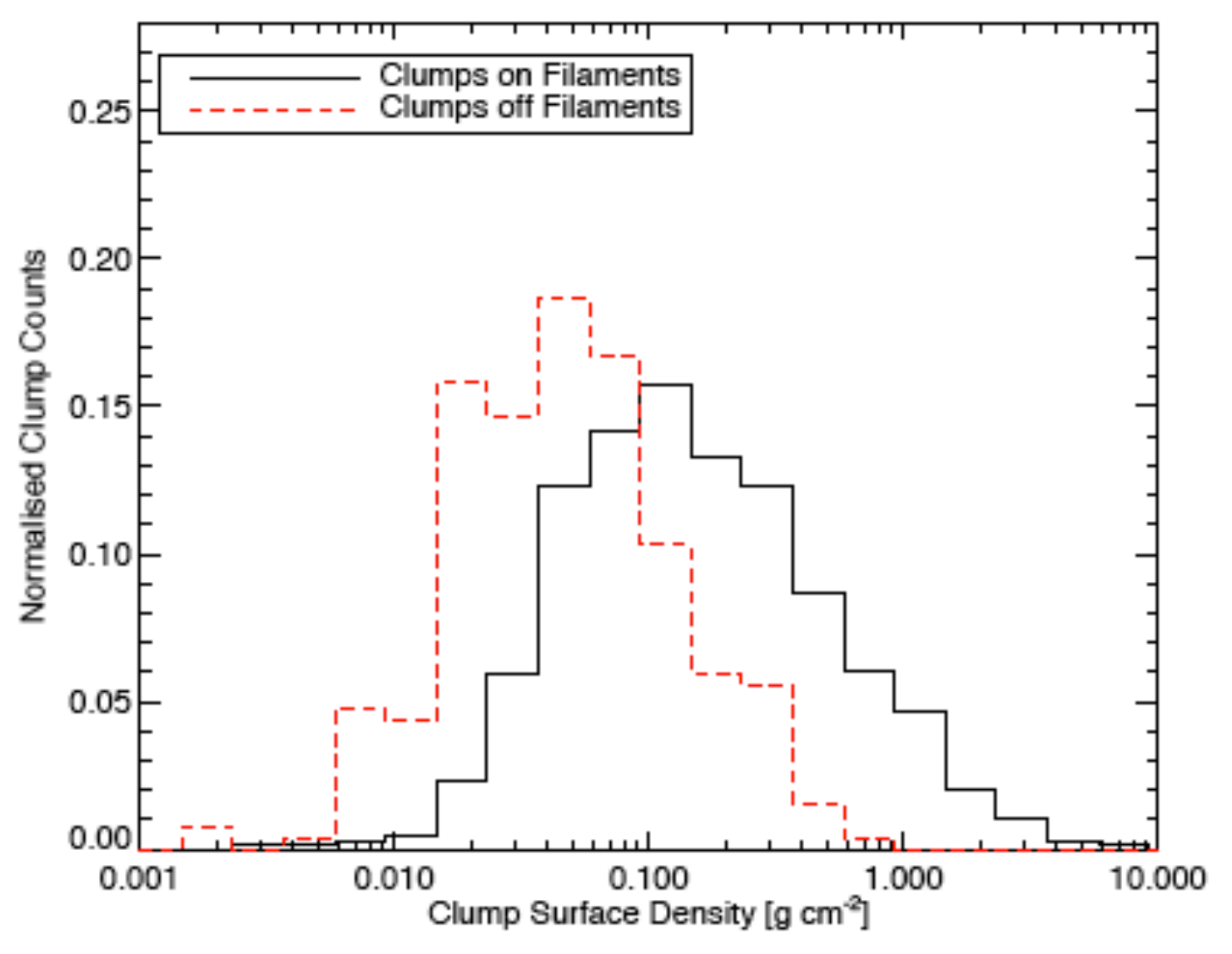}
\caption{Clump surface density distribution for clumps located within
filaments (black solid line) or outside of filaments (red dashed line) (from \citealt{Schisano+2014}).}
\label{fig:eu_fig17}
\end{figure}

Herschel-based studies by \citet{Schisano+2014} show that denser
clumps are preferentially found on filamentary structures along the
Galactic Plane (see Fig.~\ref{fig:eu_fig17}), similarly to what is
found at smaller scales in Aquila \citep{andre10} or toward Orion A
\citep{Polychroni+2013}.  Furthermore, the denser and more massive
clumps are preferentially located at the intersection of filaments
\citep{Myers2009, Schneider+12}, indicating that such extreme clumps
form from the funneling of gas along filaments.

Unbiased surveys of the Galactic Plane such as ATLASGAL, BGPS, and
Hi-GAL are revolutionizing our understanding of the formation
and star-forming evolution of dense protocluster-forming clumps.  The
ATLASGAL survey \citep{Schuller2009, Contreras2013} has been matched
to the methanol multibeam (MMB) survey \citep{Urquhart2013},
identifying 577 submm continuum clumps with maser emission in
the longitude range 280\deg $\leq l \leq 20$\deg\ (within 1\fdg5 of
the Galactic mid-plane).  94\% of methanol masers are associated with
submm dust emission and are preferentially associated with the
most massive clumps. These clumps are centrally condensed, with
envelope structures that appear to be scale-free, and the mean offset
of the maser from the peak submm column density is
0\arcsec$\pm \, 4$\arcsec.
Assuming a Kroupa IMF and a star-formation efficiency of $\sim$30\%,
they found that over two-thirds of the maser-associated clumps are
likely to form stellar clusters with masses $\geq 20$~\msol\ and that
almost all clumps satisfy the empirical mass-size criterion for
massive star formation \citep{Kauffmann+Pillai2010}.
\citet{Urquhart2013} find that the star formation efficiency is
significantly reduced in the Galactic Center region compared to the
rest of the survey area where it is broadly constant, and that there
is a significant drop in the massive star formation rate density in
the outer Galaxy. \citet{Longmore2013} compared the absolute star formation rate derived from WMAP free-free emission with the surface density of dense gas and dust
from HOPS and Hi-GAL, also deducing a much lower star formation
efficiency toward the CMZ.

A further association of ATLASGAL and UCHII regions detected by the
CORNISH survey of the inner Galactic Plane \citep{Purcell_HOPS_2012,
  Hoare2012,
  Urquhart2013b} identified a lower envelope of 0.05 g cm$^{-2}$ for
the surface density of molecular clumps hosting massive star formation
and found that the mass of the most massive embedded star is closely
correlated with the mass of the associated clump.
They find little evolution in the structure of the molecular clumps
between the methanol-maser and UCHII-region phases. The value above is significantly lower than the theoretical prediction of 1 g cm$^{-2}$ as a surface density threshold for massive star formation originally proposed by \cite{Krumholz+McKee2008}, and below the value of 0.2 g cm$^{-2}$ recently proposed by \cite{Butler+Tan2012} analysing GLIMPSE counterparts toward a sample of IRDCs.

The 1.1 mm Bolocam Galactic Plane Survey (BGPS) \citep{Aguirre2011,
  Ginsburg2013} identified over 8,500 clumps in the northern Galactic
plane \citep{Rosolowsky2010}.  BGPS observations of the CMZ can be
used to derive a clump mass spectrum for the central 500 pc 
region \citep{Bally2010_CMZ}.  Using various
assumptions about the dust temperature to derive masses,
\citet{Bally2010_CMZ} deduce clump mass spectra with a slope $\alpha$
= 2.4 to 2.7 ($dN/dM \propto M^{-\alpha}$) in the mass range $10^2$
to $10^4$~\msol, somewhat steeper than found for nearby low-mass
cores or the stellar IMF.

Motivated by the detection of the $10^5$ \msol\ CMZ cloud dubbed the
`Brick', which appears to be nearly devoid of star formation despite
its high mass and density, \citet{Ginsburg2012} searched  in a
region excluding the CMZ for massive clumps with sufficient mass to
form a young massive cluster (YMCs) with stellar mass of about
$10^4$~\msol, similar to the masses of the most massive Galactic
clusters such as NGC 3603 or Westerlund~1.  Under the assumption that
the SFE is about 30\%, \citet{Ginsburg2012} mined the BGPS clump
catalog for clumps with masses larger than $5,000$ \msol, also
requiring the radius to be smaller than 2.5 pc.  The latter criterion
was based on the arguments presented by \cite{Bressert2012} that to
form a massive, gravitationally bound cluster, the gravitational
escape speed from the cloud had to exceed the sound-speed in
photoionized plasma, about 10 \kms.  All of the 18 massive candidate
protocluster clumps were found to be actively forming massive stars or
clusters based on their IR properties.  Most of these massive clumps
are associated with well-known giant \HII\ region/star-forming
complexes such as W43, W49 and W51.  \citet{Ginsburg2012} concluded
that the duration of any quiescent ``prestellar'' or ``precluster''
phase of massive protocluster clumps (MPCCs) must have a duration less
than about 0.5 My.  Thus, massive clusters in the Galactic disk might
be assembled from smaller already actively star-forming units, rather
than arising from an isolated massive MPCC (Longmore et al., this
volume).

\begin{figure}[h!]
\includegraphics[width=0.5\textwidth]{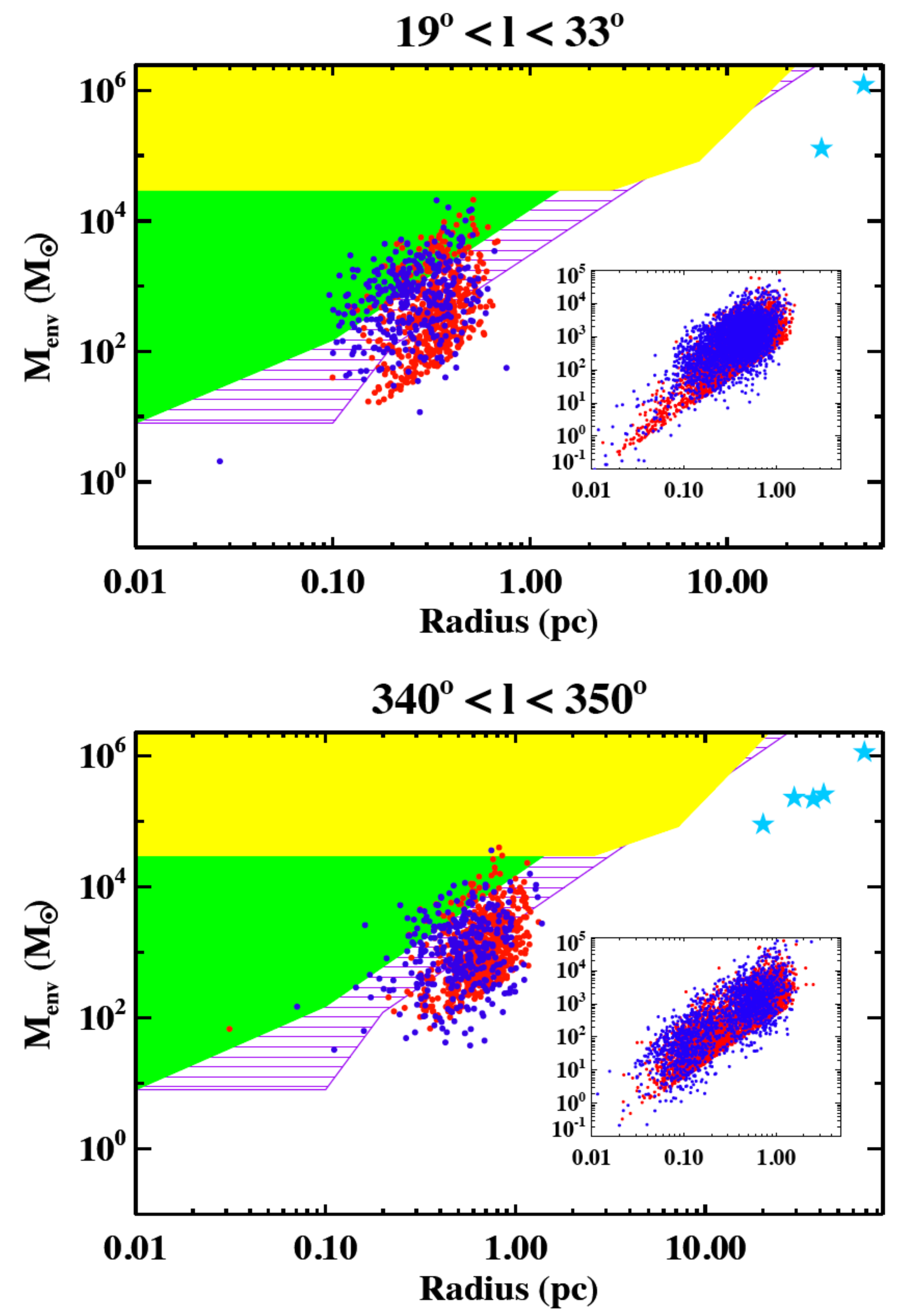}
\caption{Mass vs.\ radius for the Hi-GAL dense clumps in two ranges of Galactic longitude containing the tips of the Galactic Bar from \citet{Veneziani+2014}. Protostellar (i.e. with 70\mum\ counterpart, blue dots) and prestellar (i.e. without 70\mum\ counterpart, red dots) clumps  at the location of the tips of the Bar are reported in the main plots of the two panels. Embedded boxes show the same trend for the rest of the sample in the reported lines of sight. Light blue stars on the top right corner show where
GMC present in this area are found (see \citealt{Veneziani+2014}). See text for a description of the coloured and shaded areas.}
\label{fig:masssize}
\end{figure}

Hi-GAL will greatly increase the number of dense clumps detected.  From the first release of Hi-GAL photometric compact sources catalogues \cite{Molinari+2014a}, more than 5$\times 10^5$ individual band-merged entries can be combined for the inner Galaxy only. Downselection to sources with at least 3 detection in adiacent bands delivers an initial catalogue of nearly 100\,000 compact sources in the range +67\deg$\geq l \geq -70$\deg; for nearly 60\,000 of them a distance could be tentativley assigned, allowing to estimates sizes  ranging from 0.1 to above 1 pc and masses up to and in excess of 10$^5$~\msol\ \citep{Elia+2014b}. \emph{Dust} temperatures range from 7 to 40 K with a median $\sim
12$~K, below the value of 20 K that is often adopted to derive masses
from submm data \citep{Urquhart2013} and in better agreement
with the ammonia temperature \citep{Wienen+2012} measured toward 862
dense ATLASGAL 870~\mum\ clumps. Preliminary studies of the clump mass
functions in two very different Hi-GAL fields, toward the near tip of
the Galactic Bar and toward the Vul OB1 region ($l=59$\deg), show very
similar slopes (above the completeness limits) between 1.8 and 1.9 but
very different mass scales \citep{Olmi+2013}.
The much higher sensitivity of Hi-GAL with respect to BGPS or
ATLASGAL, and its higher spatial resolution for $\lambda \le
250$~\mum, enable larger statistical samples for clumps that may be hosting massive star formation. Fig. \ref{fig:masssize} reports the mass-radius relationship for Hi-GAL clumps limited to the lines of sight towards the tips of the Galactic Bar \citep{Veneziani+2014}. Clumps extend over areas of the plot where conditions are suited for massive star formation according to a variety of prescriptions. The green solid and the purple dashed areas are the loci with surface densities higher than 1 g cm$^{-2}$  \citep{Krumholz+McKee2008} and 0.2 g cm$^{-2}$ \citep{Butler+Tan2012}, respectively. 

\begin{figure}[t]
\includegraphics[width=0.5\textwidth]{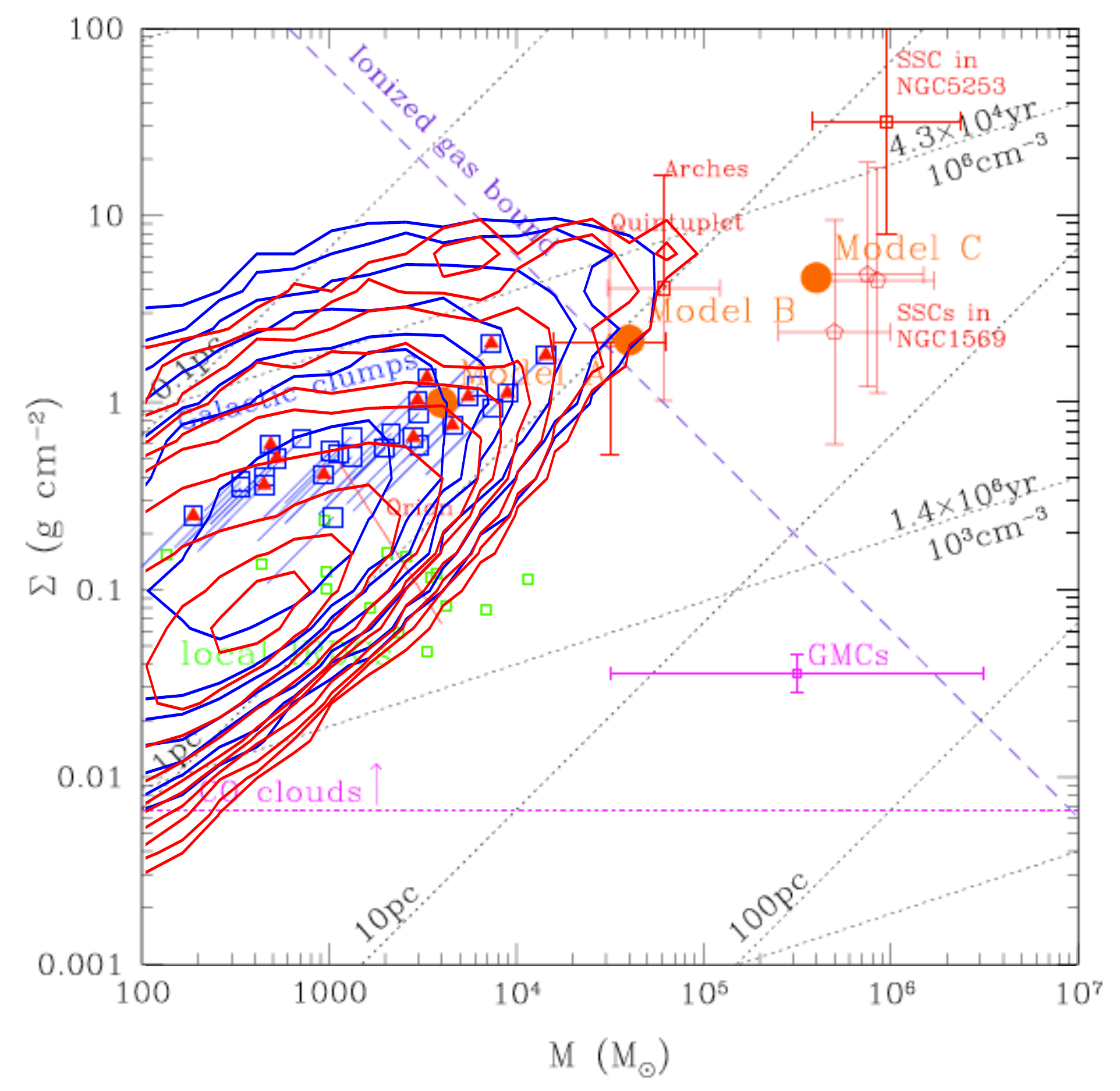}
\caption{Surface density vs.\ mass diagram from \cite{Tan2005},
overplotting the nearly 60,000 Hi-GAL detected clumps in the 67\deg\ $
\geq l \geq 19$\deg, and $-10$\deg $\leq l \leq -70$\deg, for which a
distance estimate could be made \citep{Elia+2014b}. Blue contours are for protostellar
clumps, while red contours indicate prestellar  clumps (i.e., undetected at
70 \mum). Typical locations of a variety of
structures are also indicated. Diagonal contours of constant radius and hydrogen
number density or free-fall time scale are shown with dotted
lines. The minimum surface density for CO clouds in the local Galactic
UV radiation field is shown, as are typical parameters of GMCs. The
condition for ionized gas to remain bound is indicated by the dashed
line. Locations for a selection of IRDCs (green squares) and dense
star-forming clumps (blue squares) are reported (see \citealt{Tan2005}
for a detailed description). Several massive clusters are also
indicated.
}
\label{fig:diag_tan}
\end{figure}

Larger statistical samples also allow to improve searches for potential precursors of young massive clusters, that according to \cite{Bressert2012} should populate the yellow area in fig. \ref{fig:masssize}. While none of the clumps revealed at the location of the tips of the Bar seem to fullfil this criterium, other regions of the Plane seem to contain such candidates. 
Figure~\ref{fig:diag_tan} shows surface density vs.\ mass for the
nearly 60,000 Hi-GAL clumps in the inner Galaxy for which distance estimates exist
\citep{Elia+2014b}, in the context of other structures as reported by
\cite{Tan2005}. The clumps lie in the area familiar for ``Galactic
clumps," and extend beyond it toward surface density above 5 g
cm$^{-2}$ and mass well above the few-10$^4$\msol, to the right of the dashed line marking bounded ionised gas \citep{Bressert2012},  typical of progenitors of
the most massive Galactic clusters. 
For example using the
same mass criteria as \citet{Ginsburg2012} but with a more stringent
threshold of 5 g cm$^{-2}$ in mean clump surface density, more than 100
objects are selected in Hi-GAL \citep{Pezzuto+2014}. 

\subsubsection{Theory}


Until recently, most workers considered roundish clumps and cores in
isolation as the typical sites of star formation \citep[see, e.g., the
reviews by][]{Beuther+07, diFrancesco+07, Lada+07, WardT+07}, and
modeled them as thermally-supported, equilibrium Bonnor-Ebert spheres,
perhaps undergoing oscillations.
However, since the last {\it Protostars and Planets} volume, it has
been realized that cores are highly clustered, and most of them are
embedded in filaments from which they accrete \citep{Myers2009,
  andre10, Molinari10, Arzoumanian2011, Kirk+13}.

In turbulent models of GMCs, transient density fluctuations produced by
shocks with sufficient compression and mass to exceed the Jeans mass
result in gravitationally-bound clumps and cores, while simultaneously
providing support for the clouds as a whole
\citep[see, e.g., the review by][and references therein]{BP+07}.
However, recent simulations of molecular cloud formation and evolution
including self-gravity \citep{Vazquez-Semadeni+07, Vazquez-Semadeni+09,
Vazquez-Semadeni+11, Heitsch+Hartmann08, Banerjee2009} suggest that
clouds experience global, hierarchical, and chaotic gravitational
contraction, so that no support is really necessary.  

The transformation of the diffuse ISM involves multiple stages of
collapse and fragmentation.
%
%
%
The cores in this scenario constitute the last stage of a mass cascade
\citep{Field+08} from the large, \HI\ supercloud scale, through GMCs,
down to clumps and individual protostellar cores
\citep{Clark+Bonnell05, Vazquez-Semadeni+07, Vazquez-Semadeni+09}.
Collapse occurs along the smallest dimension first, forming sheets,
then filaments, and finally clumps, as is expected for pressure-free
structures \citep{Lin1965}.
%
%
This condition applies to GMCs, because their masses are much larger
than their Jeans mass.
Clumps and cores
in this scenario are parts of dynamic rather than hydrostatic structures.

\subsection{Triggered Star Formation}

Star formation can be triggered by dynamical processes
such as protostellar outflows, FUV-driven bubbles,
expansion of \HII\ regions,
supernova remnants, superbubbles, etc.
\citep{elmegreen:1998, deharveng:2010, elmegreen:2000}.  Triggering
seems to be associated with a local increase of the star formation
rate and efficiency (SFR and SFE).  Estimates of the percentage of
YSOs formed by triggering in the Milky Way lie between 25 and 50\%
\citep{Snider+2009, deharveng:2010, thompson:2012}.

In general, any agent that results in converging flows or compression
on sufficiently large scales to overcome internal pressure or
turbulent support can create conditions for the onset of gravitational
instability.  Observations show that the highest SFRs occur in regions
of extreme pressure and density encountered in galaxy collisions and
mergers.  The processes act on different scales ranging from small
($\sim$ pc) scale compression of globules, intermediate scale ($\sim$
1 to 10 pc) triggering by expanding FUV-driven bubbles such as
NGC~2023 in Orion and EUV-powered \HII\ regions, to larger scale
triggering by expanding superbubbles.

Models of triggered star formation were motivated more than a
half-century ago by observations of age sequences in nearby OB
associations
\citep{Blaauw1991}.
For example, the 10--15 My old sub-group Ori OB1 is located in the
northwest portion of Orion.  A sequence of ever younger groups (the
1c, 1b, and 1d sub-groups, \citealt{Brown+1994}) extends toward the
southeast part of Orion toward the Orion Nebula, which marks the
location of the youngest subgroup.

Spatial segregation of age sequences prompted models of triggered,
self-propagating star formation in which young second-generation stars
form in the shock-compressed layers produced by D-type ionization
fronts propagating into clouds with speeds of a few \kms\
\citep{ElmegreenLada1977}.  If the second generation stars are
massive, their own \HII\ regions continue the process, forming third
generation stars.  However, if the clouds have large internal density
variations,
the D-type shock could lack the coherence required to collect sufficient
mass to allow formation of massive stars.

Triggering can occur through compression of pre-existing clumps and
cores by application of an external pressure.  For example, because
the pressure exerted by ionization tends to be orthogonal to the
ionization front, clouds and globules overrun by \HII\ regions can be
compressed.
In principle, application of any external pressure can lead to
collapse.  However, if the pressure jump is too large, the momentum
transfer to the denser medium being overrun causes it to be
disrupted rather than compressed.  Triggering is most likely when the
shocks associated with the external pressure have speeds below the
escape speed of the cloud.
In this scenario, second generation stars are expected to be younger
than the first generation and to have random velocities with respect
to them, because they retain the velocity dispersion of the pre-existing
cores.

In the collect and collapse mode \citep{ElmegreenLada1977}, expanding
bubbles sweep up dense shells of gas from the surrounding medium.  As
such shells accumulate and decelerate, they can become subject to
local gravitational instability and collapse resulting in star
formation.  In this scenario, the younger second generation stars are
expected to be moving away from the older generation with a speed of 1
to 10 \kms .  If the medium is relatively uniform, second-generation
stars formed earlier from the decelerating shell should have larger
velocities and so be located farther from the original
stars than those formed later, resulting in an inverted age sequence.

Triggering is an attractive theoretical idea, 
but the association of YSOs with bubble rims does not necessarily
imply that their formation was triggered.  The observed morphologies
and age sequences could have arisen from spontaneous star formation
without the action of the adjacent bubble.  The propagation velocities
of ionization fronts and shocks decrease and stall when they
encounter dense gas.  Thus, even for spontaneous, untriggered star
formation, bubbles produced by adjacent star formation will run up
against other cores and clumps, giving the impression of triggering.

Proof of triggering in the Galactic context requires precise
measurements of stellar age and proper motion.  Chronometry needs to
constrain stellar ages with a time resolution better than the age
separation; evidence for age differences using near-IR high-resolution
spectroscopy \citep{Martins+2010} are still not conclusive. Proper
motions with a precision of a few \kms\ are needed to differentiate
pre-existing star formation from triggered star formation.
For the nearby star forming complexes, the GAIA mission should soon
produce the required proper motion data

\subsubsection{Recent observational results} 

Figure~\ref{fig:W345} shows young stars and star clusters at the
periphery of the giant \HII\ regions in the W3, W4, and W5 complex.
The rims of these giant \HII\ regions are surrounded by dense cores,
pillars, and hundred of YSOs identified by Spitzer.  In the largest
and oldest portion of this complex, it has been shown that the
formation of these objects was triggered by the expansion of the 3 to
5 My old W5
\citep{Koenig2008a,Koenig2008b,Koenig2011,Deharveng+2012}.
For the highest column density, most massive and luminous young clumps
found by Herschel in W3, \citet{Rivera2013} suggest an active/dynamic
``convergent constructive feedback'' scenario to account for the
formation of a cluster with decreasing age and increasing
system/source mass toward the innermost regions and creation of an
environment suitable for the formation of a Trapezium-like system.

The Spitzer satellite has revealed many thousands of bubble-shaped
ionized regions in the Galaxy. The Milky Way Project
\citep{Simpson+2012}, combined with other surveys such as the Red MSX
Source (RMS, \citealt{urquhart:2008}), led to a statistical study of
massive star formation associated with infrared bubbles
\citep{kendrew:2012}. About 67\%$\pm$3\% of the Massive Young Stellar
Objects (MYSOs) and (ultra-)compact \hii\ regions in the RMS survey
are associated with bubbles and 22\%$\pm$2\% of massive young stars
might have formed as a result of feedback from expanding \hii\ regions
(see also \citealt{thompson:2012}).  A similar result was obtained
previously by \cite{deharveng:2010} using Spitzer-GLIMPSE,
Spitzer-MIPSGAL, and ATLASGAL surveys, indicating that about 25\% of
the bubbles might have triggered the formation of massive objects.

\citet{alexander:2013} and \cite{Kerton2013} combined Herschel images
with \HI\ and radio continuum data to probe bubbles and MYSOs at their
periphery.  Spectral energy distributions of YSOs derived from
Herschel combined with the spatial distribution of sources shows age
gradients away from the bubble center in regions such as the Rosette
\citep{schneider:2010} and N49 \citep{zavagno:2010}.



Hi-GAL has also been used to 
assess the role of triggering. 
Hi-GAL sources observed at the edges of \hii\ regions were
examined. To determine if these sources and the \hii\ regions are
physically associated, we used velocity information from different
kinematical surveys (including the MALT90 survey,
\citealt{MALT90_2011}), as well as pointed observations.  In
particular, we have produced velocity maps toward 300 \hii\ regions
(mosaicing the individual MALT90 fields for large regions).
%
Preliminary results show that the velocity fields observed around
\hii\ regions using molecular line data agree well with the velocities
obtained for the ionized gas from optical and radio recombination line
surveys, implying that much of the molecular gas with embedded sources is
associated physically with the ionized regions.
However, especially in the inner Galaxy other molecular components are
observed with different velocities along the line of sight, and our
images show that young sources are also observed toward these
components. These sources seen in projection near ionized regions
might not be physically associated with them and so their formation
would have nothing to do with the ionized region and its evolution.
Cross-correlation with Hi-GAL distance information will help in
sorting out these critical cases \citep{Zavagno+2014}.

\subsubsection{Recent theoretical developments}

Numerical models of the ionization of molecular clouds 
\citep{Mellema+06, Arthur+11, tremblin:2012a, tremblin:2012b} show
that compression and photo-erosion produce pillars, elephant trunks,
globules, and shells of swept-up, shock-compressed dense gas that can
form stars.  Radiation-hydrodynamical simulations show the formation
of pillars from the curvature fluctuations resulting from turbulence
in the dense shell.  Ionization of the lower density gas behind the
dense pillar heads produces cometary globules \citep{Haworth2012,
  Haworth2013}.  These models predict distinctive velocity fields that
can be compared to observations.  \citet{Haworth2013} developed
molecular line diagnostics of triggered star formation using synthetic
observations.  \cite{walch:2013} use SPH simulations to explore the
effects of O stars on a molecular cloud, finding that the structure of
swept-up molecular gas can be either shell-like or dominated by
pillars and globules, depending on the fractal dimension of the cloud.

\section{Star formation thresholds and rates}

\subsection{Star formation thresholds}
\subsubsection{Observational studies of local clouds}

\begin{figure}[t]
\includegraphics[width=0.485\textwidth]{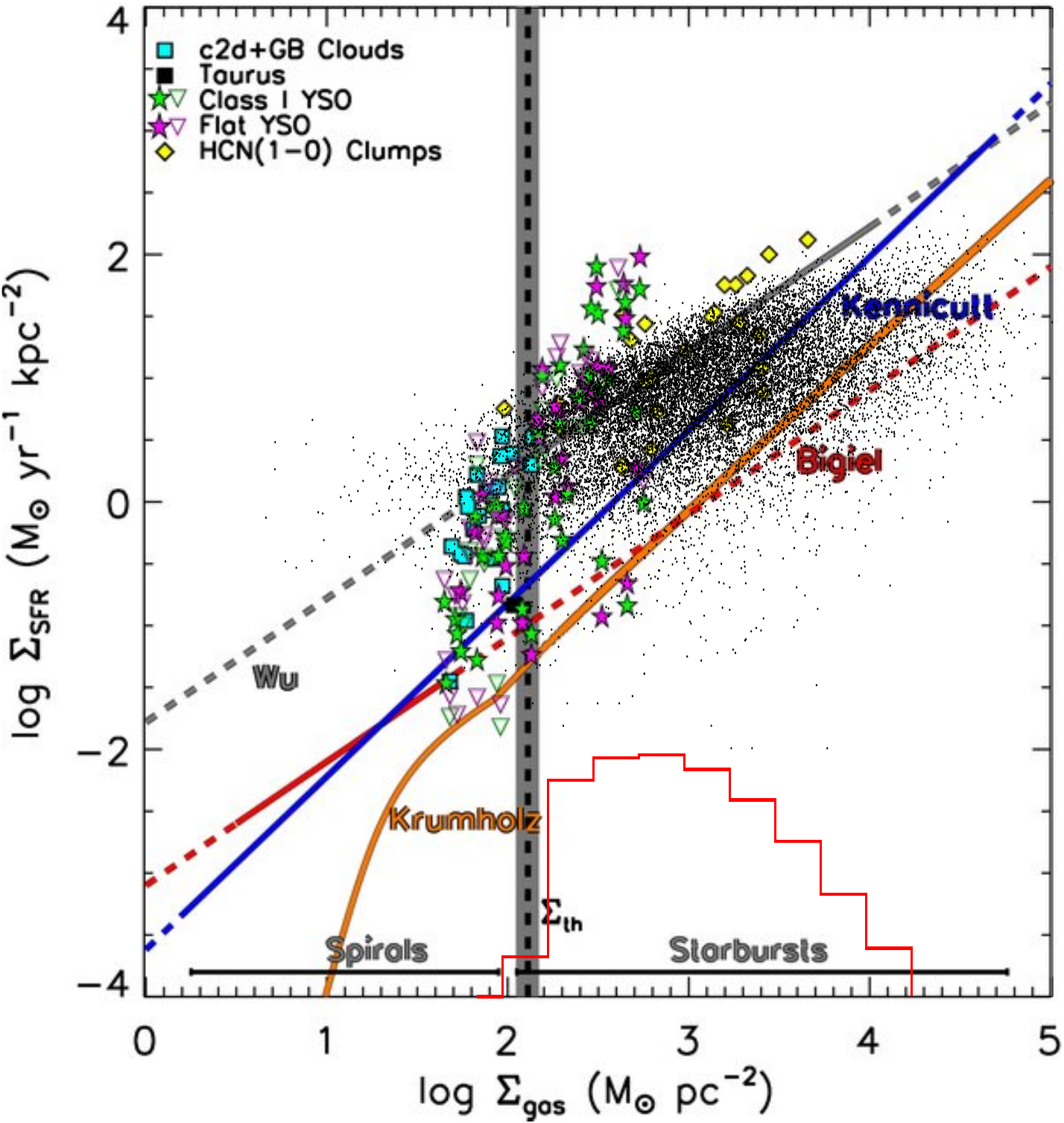}
\caption{Surface density of the SFR vs.\
mass surface density after \cite{Heiderman2010}, including data from
many studies thereby referenced.   
For comparison, the small black dots represent values derived for
individual Hi-GAL clumps \citep{Elia+2014b} classified as
protostellar based on the presence of 70 \mum\ emission.
Because no SFR estimate is available for Hi-GAL prestellar clumps, their
distribution in clump surface density is represented by the red line
histogram.
The ranges of gas surface density for spiral and circumnuclear
starburst galaxies in the \cite{Kennicutt1998} sample are denoted by
the grey horizontal lines. The grey shaded vertical band denotes the
threshold surface density $\Sigma_{th}$ of 129 $\pm$ 14\msol\
pc$^{-2}$ (see text). Several $\Sigma_{SFR} vs \Sigma_{gas}$ prescriptions are shown, and we reference to \cite{Heiderman2010} for more details.}
\label{fig:heiderman}
\end{figure}

Is there a column density threshold for star formation in local
molecular clouds? Does the SFR increase with density or column
density?  \citet{Onishi1998} mapped the Taurus molecular cloud in
C$^{18}$O and established that prestellar cores (or class 0
protostars) were found only in regions with H$_{2}$ column densities
greater than $N(H_2) \sim 8 \times 10^{21} \: {\rm cm^{-2}}$,
corresponding to a visual extinction $A_{\rm V} \sim 8.5$.  On the
other hand, \citet{Hatchell2005} found that in the Perseus cloud the
number of cores is a steeply increasing function of the C$^{18}$O
integrated intensity (and, therefore, column density).  However,
C$^{18}$O is photo-dissociated at low column densities, becomes
optically thick at high column densities, and is depleted in the
coldest, densest regions.  Thus, dust is a better tracer of the column
density.  \citet{Johnstone2004} mapped the Ophiuchus cloud at
850~$\mu$m and report that dense cores and Class-0 protostars were
found only in regions with $A_{\rm V} > 7$, indicating an extinction
threshold for the formation of dense cores.  \citet{Heiderman2010}
re-examined this issue using a sample of YSOs drawn from the Spitzer
c2d survey \citep{Evans2009} and found a steep dependence of the SFR
on the surface density, finding a column density threshold $\sim 130
\: {\rm M_{\odot}} \: {\rm pc^{-2}}$ (cf., Fig.~\ref{fig:heiderman}),
corresponding to $A_{\rm V} \sim 8.5$.
\citet{Lada2010} found that the number of YSOs is uncorrelated with
the total mass of a cloud, but correlates with the mass above a K-band
extinction of 0.8 magnitudes (corresponding to $A_{\rm V} \sim 7.3$),
again consistent with a column density threshold.

\citet{Gutermuth2011} found a steep dependence of the star formation
rate on the column density, $\Sigma_{\rm SFR} \propto \Sigma_{\rm
  gas}^{2}$, rather than a threshold and \citet{Burkert2012} also
argued that the data are more consistent with a steep, continuous
increase of the SFR surface density with gas column density.
In a first analysis of the initial Hi-GAL clump sample
\citep{Elia+2014b}, \citet{Molinari+2014b} find good agreement
with existing SFR laws (see Fig.~\ref{fig:heiderman}), but no clear
evidence for a sharp column density threshold. Clumps that are considered ``prestellar" based on the absence of 70\mum\ counterpart and which \cite{Molinari+2014b} do not use to estimate the SFR, have surface densities that are similar to the ``protostellar" clumps (the red-line histogram in Fig. \ref{fig:heiderman}). They likely represent clumps that are on the verge of forming stars.

These studies demonstrate that there is a strong dependence of the SFR
on column density, but whether/how the SFR continues to increase above
some  threshold remains uncertain.

Is there a threshold column density for the appearance of clumps and
cores on filaments?
A threshold of $A_{\rm V} \sim 8$ is reported for the appearance of
bound \emph{cores} on filaments in the Aquila region by
\citet{andre10} (see Andr\'e et al., this volume).  In the more
distant large filamentary complex in the outer Galaxy studied by
\cite{Schisano+2014} no threshold is apparent.  In particular,
Figure~\ref{fig:eu_fig18} shows a wide range of mass line densities
$M_{line}$
for filaments with bound clumps (according to Larson's criteria as applied
by \citealt{Elia+2013}).  However, the $M_{line}$ distribution for
filaments without clumps (green-line histogram) is offset to lower
values.

 

\subsubsection{Theoretical models}

Two classes of model can explain column density thresholds.  In the
first, the column density directly affects the likelihood of star
formation: star formation is strongly suppressed in gas with a column
density below the threshold value.
One plausible mechanism is photoelectric heating
of the low extinction gas by the interstellar radiation field (ISRF). In turbulent 
clouds, the complex cloud structure enables radiation to penetrate deeply into 
the clouds, and the transition from 
warm,  unshielded gas to cooler, shielded gas occurs only when the line-of-sight 
extinction exceeds $A_{\rm V} \sim$ 8. 
\citet{Clark2013} show that such a model naturally explains the
correlation between the mass of gas with $A_K > 0.8$ and the SFR found
by \citet{Lada2010}.  In their simulations gas with a line-of-sight
extinction $A_K > 0.8$ is heated only by cosmic rays, resulting in
temperatures lower than in the low-extinction regions that are heated
and partially ionized by the ISRF.  Therefore, the Jeans mass in the
high extinction regions is considerably lower, making gravitational
collapse and star formation more likely.

The second class of models is based on a dependence on the {\it
  volume} density, $n_{th}$, and minimum size, $L_{min}$.  Only
regions with column densities exceeding $n_{th} L_{min}$ can form
stars.  The existence of a volume density threshold is motivated by
observations that show that
a strong correlation exists between the amount of dense gas
and the SFR \citep[see, e.g.,][and references therein]{Kennicutt+Evans2012}.
%
%
However, simple analytical models based on this correlation tend to
predict a continuous dependence of the SFR on the gas density, albeit
perhaps steep, rather than the existence of a threshold density
\citep{Krumholz+2012, Burkert2012}.





\subsection{Star Formation Rates}

\subsubsection{Observations}

The RMS survey has provided statistically significant determinations of
the time scales and luminosity functions of massive YSOs and UCHII
regions \citep{Mottram2011, Davies2011}. Both the radio-quiet MYSO
phase, which could correspond to the swollen, cool, rapidly accreting
phase predicted by \citet{Hosokawa2010}, and the
UCHII-region phase have time scales of  one to a few times 10$^5$
years, depending on the luminosity. The luminosity functions of the two
phases are different, with UCHII regions detected up to luminosities of
$\sim$10$^6$~\lsol\ but no MYSOs above 10$^5$~\lsol, which could be due to
rapid evolution of the latter because the MYSO lifetimes become
comparable to the Kelvin-Helmholz time scale, as predicted in the
evolutionary models of \cite{Molinari+2008}.  \citet{Davies2011} found
that luminosity functions and spatial distributions are consistent with
accelerating accretion rates as the MYSO grows in mass (predicted by
turbulent core and competitive accretion  models).
Their results rule out models with constant or decreasing accretion
rates, implying a global average Galactic SFR of 1.5--2
\msol~y$^{-1}$.
For comparison, the free-fall estimate of the SFR obtained by dividing
the total molecular mass in the Galaxy by the free-fall time at the
mean density of the molecular gas, is a few $\times
10^2$~\msolyr. Thus, star formation on average in the Galaxy is highly
inefficient, with only $\sim 1$\% of the molecular gas mass being
converted to stars in a free-fall time. Equivalently, the depletion
time scale of the molecular gas is $\sim 1$~Gy in the Milky Way,
similar to values inferred for nearby galaxies \citep[see,
e.g.,][]{Leroy2013}.

\subsubsection{Theory}

Models produced in recent years to explain the observed inefficiency
of star formation can be grouped into two broad classes
%
(see also the chapters by Dobbs et al.\ and Padoan et al.\ in this
volume): ``cloud-support'' models, in which the SFR is regulated by
turbulence that globally maintains the cloud in quasi-equilibrium
conditions while inducing local compressions that cause a small
fraction of cloud's mass to collapse per global free-fall time; and
``cloud-collapse'' models, in which the clouds are assumed to be
undergoing collapse on all scales, with small-scale collapses
occurring earlier than larger-scale ones, and with an increasing SFR,
until the clouds are eventually destroyed by stellar feedback before
much of their mass is turned into stars.

A common assumption in all models is that the clouds are internally
turbulent and isothermal and therefore characterized by a lognormal
density PDF \citep{Vazquez-Semadeni1994, Padoan1997, Passot1998}.
The densest regions in the cloud have shorter free-fall times and
therefore collapse earlier if they occur in connected regions with
more than one Jeans mass. Fluctuations with sufficiently high density
will collapse on much shorter time scales than the whole cloud,
producing the ``instantaneous'' SFR within the cloud. The models
differ in the way they choose the threshold densities defining the
instantaneously-collapsing material, in the time scale associated with
this instantaneous SF, and in the global physical conditions in the
cloud.


In the cloud-support models \citep[]{Krumholz2005, Hennebelle2011,
  Padoan2011}, the clouds are in a stationary state and no time
dependence is considered.  It is interesting to calculate the fraction
of the total mass that is converted into stars over a cloud's
free-fall time, $\epsilon_{\rm ff}$ (this has been called the ``star
formation rate per free-fall time'' by \citealt{Krumholz2005},
although by definition it is an \emph{efficiency} achieved on the
free-fall time scale).
To determine $\epsilon_{\rm ff}$, the cloud is assumed to be in virial
equilibrium and to satisfy the Larson linewidth-size relationship.
The various models then differ in their choice of the characteristic
densities and associated time scales.

\citet{Krumholz2005} assume that the time scale is close to the global
cloud free-fall time, and that the density is that whose corresponding
thermal Jeans length equals the sonic scale of the turbulence. The
resulting Jeans masses are $\sim 0.1 ~ {\rm to} ~ 2 ~M_{\odot}$ (for T
= 10--40 K), comparable to the mean masses of stars.  Note, however,
that numerical simulations of isothermal, driven turbulence do not
confirm the existence of the required hypothetical clumps that are
simultaneously subsonic and super-Jeans \citep{Vazquez-Semadeni08}.
%
%
\citet{Padoan2011} instead assume that the relevant time scale is the
free-fall time at the threshold density, and that this density is
determined by the magnetic shock jump conditions and the magnetic
critical mass for collapse. Finally, \citet{Hennebelle2011} assume
that the density threshold is such that its {\it turbulent} (i.e.,
including the turbulent velocity dispersion as a source of pressure)
Jeans length is a specified fraction of the system scale, implying
that this threshold is scale-dependent because the turbulent motions
in general exhibit a scale dependence described by the turbulent
energy spectrum. For the time scale, these authors assume that each
density above the threshold collapses on its own free-fall time, a
feature they term ``multi-free-fall.'' They also propose
multi-free-fall versions of the \citet{Krumholz2005} and
\citet{Padoan2011} theories.

In general, the turbulent-support models predict the dependence of
$\epsilon_{\rm ff}$ on various parameters of the clouds, such as the
rms turbulent Mach number ${\cal M}$, the virial parameter $\alpha$
(the ratio of kinetic to gravitational energy in the cloud), and, when
magnetic fields are considered,
the magnetic $\beta$ parameter (the ratio of thermal to magnetic
pressure).
\citet{Federrath2012} show that the multi-free-fall versions of these
models all predict $\epsilon_{\rm ff} \sim 1$--10\%, in qualitative
agreement with observations.

On the other hand, cloud-collapse models \citep{Zamora2012} assume
that the clouds are in free-fall and therefore that both their density
and instantaneous star-forming mass fraction are time dependent. These
authors calibrate the threshold density for instantaneous SF by
comparing to the simulations of \citet{Vazquez-Semadeni+10}, take the
associated time scale as the free-fall time at the threshold density,
and
assume that the evolution ends when the cloud is finally
destroyed. They find that the SFR is generally increasing and compares
favorably with the evolutionary times scales for GMCs in the LMC
\citep{Kawamura:2009} and the stellar-age histograms of
\citet{Palla:2000, Palla:2002}. Moreover, \citet{Zamora2013} also find
that the time-averaged SFR as a function of cloud mass compares
favorably with the relation found by \citet{Gao2004}.

An intermediate class of models has been presented by
\citet{Krumholz2006} and \citet{Goldbaum:2011}.  They compute the
time-dependent virial balance of the clouds assuming spherical
symmetry and that the expansion of {\sc Hii} regions drives supporting
turbulence in the clouds, whereas winds produced by massive stars
escape the clouds. Their models predict that the clouds undergo an
oscillatory stage that is short for low-mass clouds and longer
(several free-fall times) for more massive clouds.

A crucial difference between the various models is the assumed effect
of stellar feedback on the clouds, which ranges from simply driving
turbulence to entirely evaporating them. Recent numerical simulations
have addressed this issue, finding that while on short time scales
($\sim 1$~My) stellar ionizing feedback indeed drives turbulence into
the clouds \citep{Gritschneder2009, Walch2012}, the final effect on
time scales $\sim 10$~My is the destruction of the star-forming site
for clouds up to $10^5$~\msol \citep{Dale+12, Dale+13, Colin+13}.


\subsection{Cloud Disruption}

The low SFE in gravitationally unbound molecular clouds is a
consequence of their dissolution in a crossing time, $L / \sigma \sim$
3 to 10 My for 10 -- 30 pc clouds obeying Larson's laws.  Bound clouds
require energy inputs to be disrupted and dissociated.  Feedback from
forming stars is the most likely agent.  For instance, feedback from
low-mass (M $< $ few \msol ) stars is dominated by their outflows
(\citealt{Li2006, Wang2010, Nakamura2011}).  Although stars of all
masses generate such flows during formation, in more massive stars
other mechanisms inject greater amounts of energy after accretion
stops.  Intermediate mass YSOs (2 $< M <$ 8 \msol) produce copious
non-ionizing radiation that heats adjacent cloud surfaces to
temperatures of $10^2$ to over $6\times 10^3$ K, raising the sound
speed to 1--6 \kms\ and dramatically increasing pressure in the heated
layer.  The resulting gas expansion can drive shocks into surrounding
gas, accelerating it.  Chaotic cloud structure results in chaotic
motions.  Late B and A stars are thus an important source of heating,
motion, and turbulence generation through the action of their
photon-dominated regions.  High mass (M $>$ 8 \msol ) stars ionize
their environments, creating $10^4$~K \HII\ regions whose pressure
accelerates surrounding gas.

The growth of H$\,${\sc ii} regions and their feedback on the
structure of molecular clouds has been studied by many different
groups \citep[see, e.g.,][]{Franco1994,Matzner2002,Walch2012}.
Small molecular clouds are efficiently disrupted by ionizing radiation
on short time scales, but the largest Galactic GMCs have escape
velocities that are significantly higher than the sound speed of
photoionized gas, and are therefore not disrupted \citep[see,
e.g.,][]{Krumholz2006,Dale+12}
 
In massive GMCs radiation pressure from absorbed or scattered photons
could be important \citep{Murray2010}.  For single scattering or
absorption, the momentum transferred to surrounding gas is $\dot P
\sim L \tau / c$ $\sim 1.3 \times 10^{29} \tau L_6$~g cm s$^{-2}$
where $\tau$ is the optical depth and $L_6 $ is the total embedded
bolometric luminosity in units of $10^6$~\lsol.  Because stars
generally form in the densest parts of clouds, this momentum transfer
contributes to gas dispersal.  The effectiveness of this form of
feedback depends on both how much radiation is produced by the stars
and what fraction of this radiation is trapped by the cloud
\citep{Krumholz2013}.  Radiation can leak out through cavities in
highly structured clouds, lowering its effectiveness as a feedback
mechanism; this is the case for most Galactic GMCs.
 
During their accretion phase, all stars produce powerful MHD-powered
bipolar outflows whose momentum injection rate is 10 to 100 times
$L/c$.  When massive stars form, their outflows are likely to clear
channels through which radiation can leak out of the parent clump,
decreasing the feedback from radiation pressure.  Models of the
effects of radiation pressure on star-forming clouds
show that in low-mass clouds radiative trapping is small and that
photoionization is more important than radiation pressure for
destroying clouds \citep{Agertz2013}.  Radiative trapping could be
larger in more massive clouds, although \citet{Krumholz2013} argue
that Rayleigh-Taylor instability of pressure-driven gas dramatically
reduces its effective optical depth (but see \citealt{Kuiper+2013}).

Feedback from outflows, FUV, ionizing radiation, and radiation
pressure produce comparable momentum injection rates.  In turn, these
are similar to momentum dissipation rates for typical parameters.
This points to a dynamic balance between dissipation and injection.
Furthermore, the low SFE in typical GMCs suggests that once the SFE
reaches 2--20\%, feedback wins and disrupts the cloud.  Failure of
feedback could lead to high SFE and formation of bound clusters
\citep{Bressert2012}.

OB stars that eventually explode as type II supernovae have lifetimes
of 3--30~My, longer than GMC lifetimes. Stellar winds, radiation
pressure, and the supernovae generated have powerful feedback on
clouds, even those at dozens or hundreds of pc from the explosion site
as observations of the local superbubbles show.  Supernovae probably
play the dominant role in driving turbulence on large scales in the
ISM \citep{MacLow2004}.

\subsection{Global mapping of thresholds and SFRs}
\label{global_map}

An assessment of the available estimates of the present day SFR in the
Milky Way was carried out by \citet{Chomiuk2011} using a common IMF
and stellar models 
applied to measured free-free emission rates and mid-IR YSO counts.
Their median result is $1.9 \pm 0.4$~\msolyr.  The GLIMPSE and MIPSGAL
surveys \citep{Robitaille2008} result in a modeled SFR $\sim 0.68 -
1.45$~\msolyr\ \citep{Robitaille2010}; the main uncertainty arises
from contamination by AGB stars. Using the classic method of
measuring the Lyman continuum luminosity from HII regions free-free radio continuum \citet{Murray+Rahman2010} use the WMAP data to find SFR = 1.3$\pm$0.2 \msolyr, for a Salpeter
IMF, and values between 0.9 and 2.2 \msolyr\ for a range of IMF slopes. This may be a
lower bound since it not clear that the analysis of WMAP free-free
emission properly accounts for Lyman continuum lost from the Galaxy that leaks out of
chimneys to illuminate the Lockman layer.


\citet{Kennicutt+Evans2012} present a review connecting the Galactic
star formation with extragalactic systems.  \cite{Evans2009} and
\cite{Lada2010} found that for a given surface density in nearby
molecular clouds the extragalactic Schmidt-Kennicutt relationship
underestimates the SFR, because studies of the SFR in Galactic GMCs
tend to ignore the surface density from extensive, non-star-forming
gas;
when this is included, the Galactic SFRs fit the extragalactic
relationship.  An example of this is shown in
Figure~\ref{fig:heiderman}, where \cite{Heiderman2010} plot local
clouds and Galactic clumps onto the extragalactic star formation
relationship.
The figure shows an order of magnitude range in SFR at any given gas
surface density.
For individual protostellar clumps cataloged by \cite{Elia+2014b}
using Hi-GAL data, we can derive the SFR and mass per unit area. These
are overplotted as small black dots in Figure~\ref{fig:heiderman}.

{\it Herschel} provides a wealth of data on the formation rate of
massive stars in different Galactic environments.  For example, in
four square degrees at longitudes $l =30$\deg (near the Scutum arm
tangent) and at $l=59$\deg\ \citep{russeil:2011,Veneziani+2013} Hi-GAL
found average rates of $\sim 10^{-3}$ and $\sim 10^{-4}~$
\msolyr, respectively.  In the rich massive star-forming region G305
\citet{Faimali2012} found rates of 0.01--0.02~\msolyr.  In the $l =
217-224$\deg~region situated in the outer Galaxy at Galactocentric distances greater than 10kpc \cite{Elia+2013} found a value of $\sim 5\times 10^{-4}$~\msolyr.  \citet{Veneziani+2014} analyze the star formation content at the tips of the Galactic Bar using the same method as in their 2013's paper, but also incorporating a statistical correction to predicted luminosities to account for the fact that these clumps host protoclusters rather than single massive forming stars; they find values $\sim 1.5\times 10^{-3}$ \msolyr kpc$^{-2}$, at both sides of the Bar. Globally combining all of the massive objects in the preliminary Hi-GAL catalog, limited to the +67\deg$\geq l \geq -70$\deg\ range and \textit{excluding} the inner $\sim$25 degrees of the Central Molecular Zone, \citet{Molinari+2014b} find a total SFR $\sim 0.2$~\msolyr, that can be considered a lower limit especially because only a fraction, although significant, of the Galaxy is considered. 

The conversion efficiency of molecular gas into dense potentially
star-forming clumps and into stellar luminosity has been examined by
\cite{Eden2012} and \cite{Moore2012} as a function of position, both
in and out of spiral-arms.  There is little or no increase in SFE
associated with the Scutum-tangent region and the near end of the
Galactic bar, which includes the W43 complex. 
Investigation of the Galactic distribution of the $\sim1650$ embedded,
young massive objects in the RMS survey, one third of which are above
the completeness limit of $\sim 2 \times 10^4$~\lsol
\citep{Urquhart2013}, shows that the distribution of massive star
formation in the Milky Way is correlated with a model of four spiral
arms.  The increase in SFR density in spiral arms is due to the
pile-up of clumps, rather than to a higher specific rate of conversion
of clumps into stars (a higher SFE).
That the source and luminosity surface densities of RMS sources are
correlated with the surface density of the molecular gas, suggests
that the massive star-formation rate per unit molecular mass is
approximately constant across the Galaxy. The total luminosity of the
embedded massive-star population is found to be 
$\sim 0.8 \times
10^8$~\lsol, 30\% of which is associated with the ten most active star
forming complexes.

Given the enormous reservoir of dense H$_{2}$ in the CMZ, a high SFR
might be expected there.  Local relationships \citep{Lada2012,
  Krumholz2012} extrapolated to Galactic Center conditions predict
0.18--0.74~\msolyr\ (based on extinction) or 0.14--0.4~\msolyr\ (based
on volume density).  Surprisingly, however, \citet{Yusef-Zadeh2009}
and \citet{Immer2012} found 0.08--0.14~\msolyr, also consistent with
the Schmidt-Kennicutt relationship.  Using WMAP free-free emission as
a tracer of star formation within 500~pc of the Galactic Center (about
1~kpc$^{-2}$), \citet{Longmore2013} found even lower values,
$3.6\times 10^{-3}$ to $1.6\times 10^{-2}$~\msolyr, again in
disagreement with predictions.

\section{Conclusions}

The avalanche of new infrared to radio continuum and spectroscopic
imaging of the Galactic Plane is a goldmine that we are just starting
to exploit, and yet more are desirable.
Current quadrant-scale CO spectroscopic surveys of the 3$^{rd}$ and 4$^{th}$ quadrants only deliver super-arcminute
resolution. Sub-arcmin CO surveys are required; this effort is indeed starting to be assembled (see Table 1) but more is needed.  
Additional high spatial resolution surveys in molecular tracers like NH$_3$ and other high-density tracers are also needed.
Considerable potential also resides in post-Herschel unbiased
arc-second resolution continuum surveys in the far-IR and submm, as
well as in the radio where sensitivities of current unbiased surveys are
still not adequate for detailed Galactic studies.  These will be
within reach of future facilities like SPICA, CCAT, MILLIMETRON and the SKA
pathfinders, while ALMA and JWST will be able to concentrate on
selected samples of star-forming complexes.
Large-scale sub-arcminute resolution polarization surveys in the
far-IR and submm are badly needed to ascertain the role of magnetic
fields throughout the cloud-filament-clump-core evolutionary sequence.

In our opinion, however, the real challenge and opportunities lie in
an ``industrial revolution'' of the science analysis methodologies in
Galactic astronomy.
Current approaches are largely inadequate to execute a timely and
effective scientific exploitation of complete Galactic Plane surveys
covering several hundreds of square degrees.
Progress is needed in tools 
for pattern recognition and extraction of clumps, filaments, bubbles, and
sheets in highly variable background conditions,
for robust construction of SEDs,
for evolutionary classification of compact clumps and cores,
and for 3D vector modeling of Galactic rotation incorporating streaming
motions, to name a few.
Toward building a Galactic plane knowledge base, we envision these
technical developments integrated within a framework of data mining,
machine learning, and interactive 3D visualization, so as to transfuse
the astronomer's know-how into a set of automated and supervised tools
with decision-making capabilities.  



\end{document}